\documentclass[a4paper,fleqn]{cas-sc}

\usepackage[authoryear]{natbib}
\graphicspath{{figs/}}
\usepackage{dcolumn}

\usepackage{dirtytalk}
\MakeRobust{\say}
\usepackage[version=4]{mhchem}
\usepackage{xcolor}
\usepackage[french,english]{babel}
\usepackage[autostyle]{csquotes}
\usepackage{numprint}
\usepackage{siunitx}
\usepackage{caption}
\usepackage{subcaption}
\usepackage{gensymb}
\usepackage{microtype}

\def\tsc#1{\csdef{#1}{\textsc{\lowercase{#1}}\xspace}}
\tsc{WGM}
\tsc{QE}

\makeatletter\def\Hy@Warning#1{}\makeatother

\begin{document}
\let\WriteBookmarks\relax
\def\floatpagepagefraction{1}
\def\textpagefraction{.001}

\shorttitle{}
\shortauthors{A. Borner et~al.}

\title[mode=title]{Noble Gas Fractionation Predictions for High Speed Sampling in the Upper Atmosphere of Venus}

\author[1]{Arnaud Borner}[orcid=0009-0008-0984-2542]
\cormark[1]
\ead{arnaud.p.borner@nasa.gov}
\credit{Writing - review \& editing, Writing - original draft, Visualization, Software, Methodology, Investigation, Formal analysis, Data curation, Conceptualization}
\affiliation[1]{organization={Analytical Mechanics Associates, Inc. at NASA Ames Research Center},
            city={Moffett Field},
            postcode={94035}, 
            state={CA},
            country={USA}}
       
\author[2]{Michael A. Gallis}[orcid=0000-0002-4985-6956]
\credit{Writing - review \& editing, Software, Methodology, Investigation, Data curation}
\affiliation[2]{organization={Sandia National Laboratories},
                city={Albuquerque},
                postcode={87185},
                state={NM},
                country={USA}}
                
\author[3]{Rita Parai}[orcid=0000-0002-9754-7349]
\credit{Writing - review \& editing, Visualization, Investigation, Conceptualization}
\affiliation[3]{organization={McDonnell Center for the Space Sciences, Washington University in St. Louis},
                city={St. Louis},
                postcode={63130},
                state={MO},
                country={USA}}

\author[4,5]{Guillaume Avice}[orcid=0000-0003-0962-0049]
\credit{Writing - review \& editing, Visualization, Investigation, Conceptualization}
\affiliation[4]{organization={Universit{\'e} Paris Cit{\'e}, Institut de physique du globe de Paris, CNRS},
                city={Paris},
                postcode={F-75005},
                country={France}}
\affiliation[5]{organization={California Institute of Technology, GPS Division},
                postcode={91125},
                state={CA},
                country={USA}}
                
\author[6]{Mihail P. Petkov}[orcid=0000-0002-2739-6928]
\credit{Writing - review \& editing, Visualization, Investigation, Conceptualization}
\affiliation[6]{organization={SuprAEther, LLC},
                addressline={3533 Henrietta Ave},
                city={La Crescenta},
                postcode={91109},
                state={CA},
                country={USA}}
                
\author[1]{Krishnan Swaminathan-Gopalan}[orcid=0000-0003-2537-9320]
\credit{Writing - review \& editing, Investigation}

\author[7]{Christophe Sotin}[orcid=0000-0003-3947-1072]
\credit{Writing - review \& editing, Investigation, Conceptualization}
\affiliation[7]{organization={Laboratoire de Plan{\'e}tologie et G{\'e}osciences, Nantes Universit{\'e}, Universit{\'e} Angers, Le Mans Universit{\'e}, CNRS, UMR 6112},
                city={Nantes},
                postcode={F-44000},
                country={France}}

\author[8]{Jason Rabinovitch}[orcid=0000-0002-1914-7964]
\credit{Writing - review \& editing, Writing - original draft, Visualization, Methodology, Investigation, Funding Acquisition, Formal analysis, Conceptualization}
\affiliation[8]{organization={Stevens Institute of Technology},
            city={Hoboken},
            postcode={07030}, 
            state={NJ},
            country={USA}}

\cortext[1]{Corresponding author}


\begin{abstract}
  Venus, our neighboring planet, is an open-air laboratory that can be used to study why Earth and Venus evolved in such different ways and even to better understand exoplanets. Noble gases in planetary atmospheres are tracers of their geophysical evolution, and measuring the elemental and isotopic composition of noble gases in the Venus atmosphere informs us about the origin and evolution of the entire planet. In this work we describe a new mission concept, Venus ATMOSpheric - Sample Return (VATMOS-SR), that would return gas samples from the upper atmosphere of Venus to Earth for scientific analysis. This could be the first sample return mission for an extraterrestrial atmosphere. To ensure it is possible to relate the composition of the sampled gases (acquired when the spacecraft is traveling $>10$~km/s) to the free stream atmospheric composition, large-scale numerical simulations are employed to model the flow into and through the sampling system. In particular, an emphasis is placed on quantifying noble gas elemental and isotopic fractionation that occurs during the sample acquisition and transfer process, to determine how measured isotopic ratios of noble gases in the sample would compare to the actual isotopic ratios in the Venusian atmosphere. We find that lighter noble gases are depleted after they are sampled compared to the freestream conditions, and heavier ones are enriched, due to the high pressure gradients present in the flowfield. We also observe that lighter noble gases are more affected than heavier ones by changes in the freestream conditions. Finally, we observe that, in general, the numerical parameters do not have a major impact on the observed fractionation. We do, however, note that the freestream velocity and density have a major impact on fractionation, and do need to be precisely known to properly reconstruct the fractionation in the sampling system. We demonstrate that the sample fractionation can be predicted with numerical simulations, and believe that VATMOS-SR, which could be the first mission to bring back samples from another planet, could answer key scientific questions related to understanding the evolution of Venus.
\end{abstract}



\begin{keywords}
Venus \sep Atmospheres, composition \sep Abundances, atmospheres \sep Planetary formation
\end{keywords}

\maketitle

\section{Introduction}
\label{sec:introduction}

Noble gases in planetary atmospheres are tracers of the geophysical evolution of the host planet. They carry the fingerprints of processes driving atmospheric composition, including: the original supply of volatiles from the solar nebula, delivery of volatiles by asteroids and comets, rates of escape of planetary atmospheres, degassing of the interior, and its timing in the planet's history~\citep{ref:Pepin1991,ref:chassefiere2012}. To date, planetary scientists have successfully made measurements of noble gases in gaseous form at Earth, Mars, Jupiter, and comet 67P/Churyumov-Gerasimenko. An incomplete set of noble gases (Ar and Ne) were measured in Titan’s atmosphere by the Huygens probe. Noble gases contained in the atmospheres of Saturn, Venus, and the ice giants, Uranus, and Neptune, have been measured only in part or remain unmeasured, leaving major gaps in our understanding of the distribution of volatile elements in the solar system. In order to be able to understand the origin(s) of terrestrial planets and to compare their evolution, a major observational missing link in our understanding of Venus' evolution are the elemental and isotopic compositions of noble gases and stable isotopes in its atmosphere, which remain poorly known~\citep{ref:chassefiere2012}. The concentrations of heavy noble gases (Kr, Xe) and their isotopes are mostly unknown, and our knowledge of light noble gases (He, Ne, Ar) is incomplete and imprecise \citep{avice_noble_2022}. NASA's community-based forum, the Venus Exploration Analysis Group (VEXAG), has placed a high priority on obtaining such measurements in its \say{Goals, Objectives, and Investigations} document \citep{ref:VEXAG2019}. A sample return mission will be able to provide unambiguous scientific data to answer the science questions regarding the origin and evolution of Venus. Three high-impact science objectives for the exploration of the Venus atmosphere are presented in Table~\ref{fig:science-goals}. Of note is that, while the ratio of $^{15}$\ce{N} to $^{14}$\ce{N} is an important science objective, it is out of the scope of this work to study the fractionation of different isotopes of a relatively reactive non-noble gas, and only the fractionation of noble gases is studied. Also important is the fact that the listed scientific objectives generally require a precision of approximately 1\% of the measured value for isotopic ratio measurements.

\begin{table*}
\centering
\begin{tabular}{|>{\centering\arraybackslash}m{1cm}|>{\centering\arraybackslash}m{4.5cm}|>{\centering\arraybackslash}m{3.5cm}|>{\centering\arraybackslash}m{4.5cm}|}
\hline
\multirow{4}{*}{\rotatebox[origin=c]{90}{\parbox{4.1cm}{\centering \textbf{Science Goal:} \\ Understand Venus' evolution}}} & \textbf{Science Objectives} & \textbf{Physical Parameter} & \textbf{Observables (precision)} \\
\cline{2-4} 
& Determine if Venus’ isotopic composition falls on the primordial solar composition & Isotope composition of noble gases and nitrogen & Isotope ratios: $^{132}$Xe/$^{130}$Xe (1\%), $^{136}$Xe/$^{130}$Xe (1.5\%), $^{15}$N/$^{14}$N (1\%) \\
\cline{2-4} 
& Distinguish among models for planetary-scale volcanism & Elemental ratios and isotopic compositions of noble gases & Relative abundance of He  ($^4$He/$^{40}$Ar) and isotopic ratios $^3$He/$^4$He (50\%), $^{129,131-136}$Xe/$^{130}$Xe (1\%) \\
\cline{2-4} 
& Determine whether escape mechanisms were more active on Venus than on Earth & Fractionation of Xe & Amount of $^{129}$Xe, $^{124}$Xe/$^{130}$Xe, $^{126}$Xe/$^{130}$Xe, $^{128}$Xe/$^{130}$Xe (1–2\%) \\
\hline
\end{tabular}
\caption{Three high-impact science objectives that are enabled by measuring isotope ratios in Venus' atmosphere.}
\label{fig:science-goals}
\end{table*}

VATMOS-SR (Venus ATMOSpheric - Sample Return; a new mission concept that evolved from the \say{Cupid's Arrow} and \say{Cupid's Boomerang} SmallSat mission concepts) is a mission concept that would return a gas sample from the upper atmosphere of Venus to Earth for scientific analysis. Previous iterations of the mission concept required the spacecraft to make in-situ measurements using an onboard mass spectrometer~\citep{ref:sotin2018agu,ref:sotin2019,ref:rabinovitch2018,ref:rabinovitch2019a}, though the current mission design focuses on returning the Venus atmospheric samples to Earth to leverage large, advanced terrestrial measurement facilities that are not possible to integrate into a SmallSat. To ensure that an atmospheric measurement is as representative of the atmosphere as possible, it is a requirement to acquire a gas sample below an altitude referred to as the homopause. Below the homopause, atmospheric gases are presumed to be well-mixed (no spatial gradients), and samples are representative of their naturally occurring concentrations. The homopause at Venus is expected to be at an altitude of $\approx120$~km~\citep{ref:mahieux2012,ref:mahieux2015}. Even if sampling occurs beneath the homopause altitude, the possibility of sample fractionation due to the proposed sampling methodology must also be considered. In this work, the term fractionation refers to the separation of different gases from a homogeneous mixture, and we investigate whether the noble gases collected in the sampling system would have a relative composition that is the same (no fractionation) or different (fractionation) from the freestream (Venusian atmosphere).

As such, the present study focuses on determining whether atmospheric noble gases collected in the upper atmosphere of Venus ($\approx110$~km altitude) with a spacecraft traveling at $>10$~km/s are representative of the freestream composition. Or, if the composition does change between the acquired gas sample and the Venus atmosphere, if these changes can be predicted with numerical simulations, as this would also be acceptable from a mission standpoint. The formation of a merged shock-boundary layer (compression layer) in front of a vehicle traveling at hypersonic speeds under rarefied conditions adds complexity to the sampling process. Shock layers create large density gradients in the flow which may alter the composition of the gas through diffusion. Due to the challenges associated with reproducing VATMOS-SR expected flight conditions in ground testing facilities, it is necessary to rely on high-fidelity numerical simulations to shed some light on this complex problem.
\section{Background}
\label{sec:missionconcept}

\begin{figure}
 \centering
 \includegraphics[width=0.6\textwidth, angle=-90]{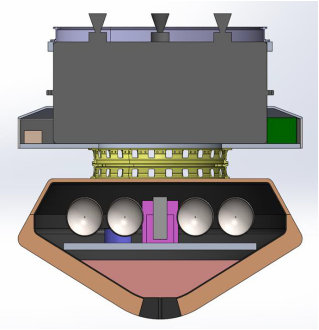}
 \caption{Possible design for the VATMOS-SR Vehicle, composed of a cruise stage, separation rings, and a probe (total diameter less than 1~m). The four sampling tanks are located in the probe. The predicted dry mass is less than 75~kg (100~kg with 30\,\% contingency). A 60$^\circ$ sphere-cone aeroshell is shown, with the sampling system inlet located at the stagnation point of the vehicle. However, the design later evolved to a 45$^\circ$ sphere-cone, which is the geometry used in the simulations presented in this work, for aerodynamic stability.}
 \label{fig:spacecraft}
\end{figure} 

\begin{figure}
 \centering
 \includegraphics[width=0.6\textwidth]{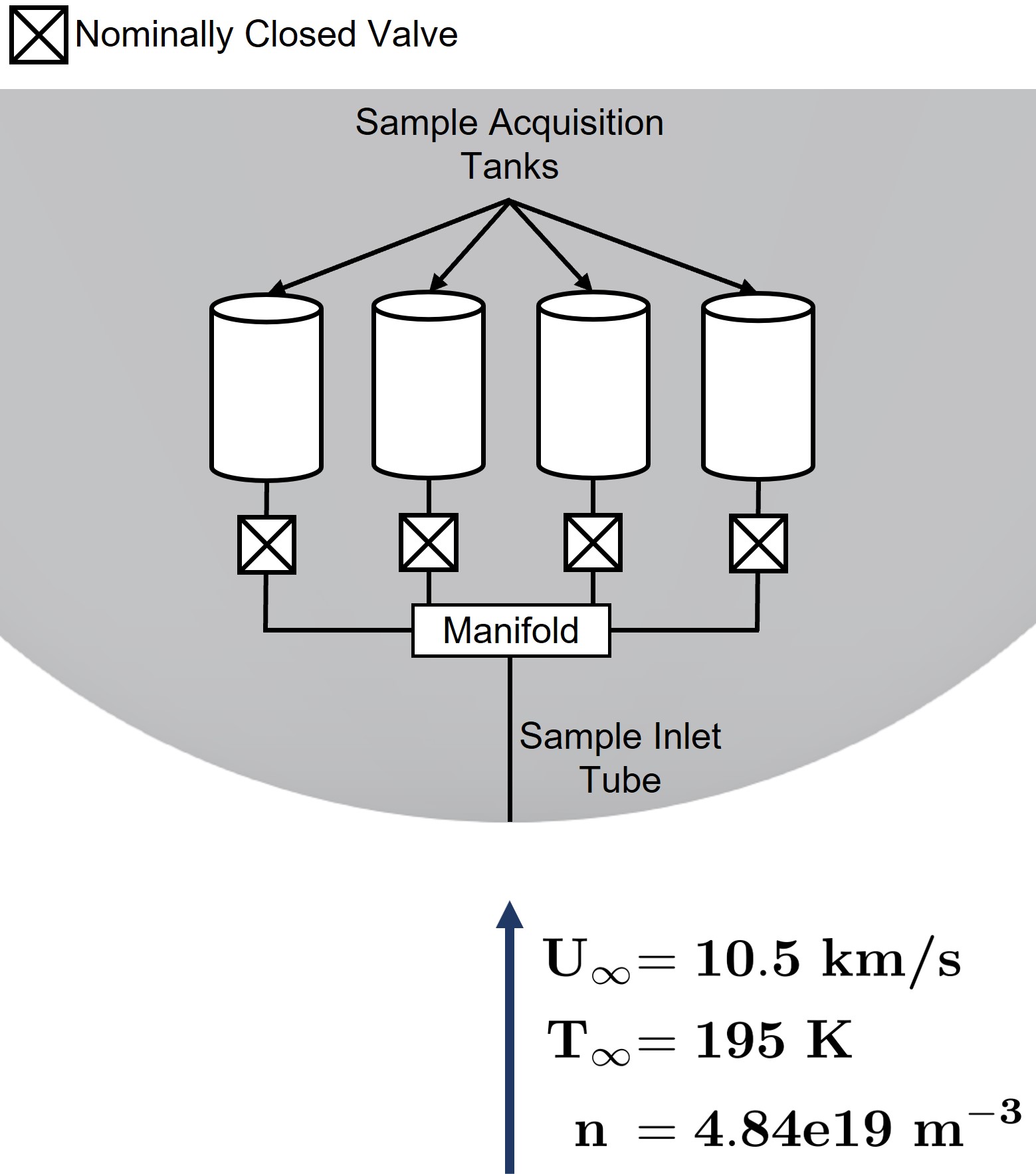}
 \caption{Schematic of the sampling system for VATMOS-SR, including the nominally closed Mindrum valves.}
 \label{fig:sampling-schematic}
\end{figure}

If flown and successful, VATMOS-SR could return the first atmospheric samples from another planet, and these samples would be analyzed in state-of-the-art laboratories on the Earth. VATMOS-SR evolved from the Cupid’s Arrow mission concept~\citep{ref:sotin2018agu}, where  a small satellite was originally proposed to make in-situ measurements in Venus’ atmosphere.
The VATMOS-SR spacecraft (Fig.~\ref{fig:spacecraft}) is an atmospheric skimmer, $\approx80$~cm in diameter, that would sample the Venus atmosphere below the homopause. Four samples of the Venusian atmosphere would be acquired at periapsis, and then the samples would be \emph{returned to Earth}, a major change from earlier iterations of the mission concept, for laboratory analysis. The velocity at periapsis, where sampling is to occur, is expected to be $\approx10.5$~km/s, and the altitude is expected to be $\approx110$~km.
The sample collection system will consist of a gas inlet tube leading to a manifold which is connected to four separate storage tanks. A non-ablating thermal protection system (TPS) will be used on the nose of the vehicle to minimize the possibility of sample contamination. The gas flow into each sample acquisition tank will be individually controlled by microvalves, likely similar to the ones conceived by Mindrum (henceforth referred to as \say{Mindrum valves}); the Mindrum valves have flown on multiple space missions in the past (\emph{e.g.}, included in the Sample Analysis at Mars (SAM) Investigation and Instrument Suite~\citep{ref:Mahaffy2012}). Fig.~\ref{fig:sampling-schematic} shows a simplified schematic of the sampling system used in this study. In the design that we used for this study, the sample inlet tube has a diameter of 8~mm, and a length of 100~mm before the first 90$^\circ$ bend, followed by four shorter tubes of diameter 6~mm and length 12.5~mm that are quarter symmetric. A Mindrum valve follows each of those shorter tubes. After each valve, there are an additional two tubes, with the final one leading to a tank of volume 1~L. There are a total of four sampling tanks in the system.

As previously stated, one potential challenge with this mission architecture is whether the sample that is collected by the spacecraft is representative (in terms of composition) of Venus' atmosphere. Therefore, it should be demonstrated that elemental and/or isotopic fractionation processes do not significantly alter the \emph{relative} concentrations of the gas samples that the spacecraft would acquire, or that any alterations in gas composition can be predicted. Due to the high velocity of the spacecraft, high altitude, and associated high enthalpy of the flow, a non-equilibrium rarefied flow field is expected around the spacecraft. Furthermore, a variety of different length scales are relevant for this problem, ranging from the overall spacecraft diameter ($\approx80$~cm) to flow-path constrictions ($\approx2$~mm). Moreover, significant variations in density between the freestream conditions and post-shock flow through the sampling system generate extremely stringent computational requirements to run these simulations in a reasonable time frame. For this particular problem, due to the rarefied nature of the flow, we estimate the shock thickness to be on the order of centimeters. Based on the flow regime, the direct simulation Monte Carlo (DSMC) methodology is chosen for this work. Moreover, even for low Knudsen number flows, the gradients of this particular problem are such that a kinetic theory approach is necessary. DSMC has been shown to provide solutions in agreement with the Boltzmann equation for both non-continuum and highly non-equilibrium flows~\citep{wagner1992convergence,gallis2004molecular}.
Simulations are performed using the Stochastic PArallel Rarefied-gas Time-accurate Analyzer (SPARTA), an open-source software package developed by Sandia National Laboratories~\citep{plimpton2019direct}.
The objective of this work is to demonstrate the capability to model the performance of the sampling system with a flight-relevant geometry for the VATMOS-SR Mission Concept, with an emphasis on quantifying any elemental or isotopic fractionation that occurs for the noble gas species of interest.
\section{Venus Atmosphere}
\label{sec:venusatmosphere}
The current state-of-the-art regarding the elemental and isotopic compositions of noble gases in the Venus atmosphere has been reviewed recently (see~\citep{avice_noble_2022} and references therein). Most data have been collected by American and Soviet space missions operated during the 1970s and 1980s. Overall, those data point toward relatively high abundances of neon, argon, krypton, and xenon in Venus' atmosphere when compared to Earth and Mars. This suggests that Venus may have inherited more volatile elements than the two other planets. These volatile elements could have been present in Venus' building blocks or could have been delivered later on in its evolution by volatile-rich bodies such as carbonaceous asteroids and/or comets. The isotopic composition of neon points towards a significant contribution from solar-derived neon, although the lack of krypton and xenon isotope data do not allow one to test this hypothesis of volatile origins for Venus. Noble gas isotopes are also useful tracers of a planet's geodynamics over its entire geological history. The relatively low $^{40}$Ar/$^{36}$Ar ratio ($\approx1$, low compared to Earth's atmospheric ratio of $\approx300$) suggests that Venus outgassed less of its radiogenic $^{40}$Ar compared to the case of Earth~\citep{kaula_constraints_1999}, although outgassing histories are model-dependent \citep{orourke_thermal_2015} and rely on estimates of the K/U ratio for the bulk planet (see the discussion by \citep{gillmann_long-term_2022}). The fact that Venus' atmospheric xenon does not present an extremely high $^{129}$Xe/$^{132}$Xe ratio~\citep{avice_noble_2022} suggests that, contrary to the case of Mars, Venus did not suffer from a massive escape of its primordial xenon followed by the degassing of radiogenic $^{129}$Xe produced in the silicate portions of the planet during the first 150 Myr of its history by radioactive decay of now extinct $^{129}$Xe \citep{swindle_martian_2002}. Again, only a few measurements with large uncertainties are available, but the selected future mission DAVINCI \citep{garvin2022} should provide additional data measured \textit{in situ} with improved precision to shed new light on planetary volatile origins and geodynamics. 

The precision and accuracy achievable by \textit{in situ} measurement is limited by considerations of instrument mass, volume, power consumption, and analysis time during transit through the atmosphere. To mitigate these challenges, Venus atmospheric sample return and analysis in terrestrial laboratories is critical. As the expected atmospheric properties at the sampling altitude will provide the initial conditions for the DSMC simulations performed in this work, it is also important to understand the bulk atmospheric composition. The bulk Venus atmospheric composition is dominated by two species: the atmosphere is $96.5\pm 0.8\,\%$ \ce{CO2} and $3.5 \pm 0.8\,\%$ \ce{N2} \citep{Fegley2014}, by volume. Other species are present at trace abundances that sum to less than $0.1\,\%$ of the atmosphere by volume: \ce{SO2}, \ce{Ar}, \ce{H2O}, \ce{CO}, \ce{He}, \ce{Ne},  \ce{H2SO4}, \ce{OCS}, and \ce{H2S} are observed in the few ppmv to $\approx150$~ppmv level, while \ce{O3}, \ce{HCl}, \ce{HF}, \ce{SO}, \ce{NO}, \ce{SO3}, \ce{CS}, \ce{CS2}, \ce{Kr}, and \ce{Xe} are present at the few ppbv level \citep{ref:Arney2014, Fegley2014, ref:Oschlisniok2021, avice_noble_2022, ref:Mahieux2023, ref:Mahieux2023b}. Droplets of condensed \ce{H2SO4} and \ce{H2O} \citep{krasnopolskypollack1994} form a cloud layer from $\approx48$ to $70$~km above the surface \citep{titovetal2018}. Importantly, the composition of the atmosphere has been observed to be variable in space (altitude, latitude, and longitude) and time \citep{marcqetal2018}. In particular, the concentrations of \ce{H2O}, \ce{SO2}, and \ce{CO} appear to be altitude-dependent, with distinct vertical profiles above and below the cloud layer \citep{marcqetal2018}. 

The homopause is the altitude region where a transition in the dominant mode of gas transport occurs~\citep{vonzahnZ79}. Below the homopause, gas transport is dominated by eddy diffusion and inert gases are well-mixed; their concentrations vary with altitude with the same scale height. Above the homopause, gas transport is dominated by molecular diffusion, which separates gas species by altitude based on their molecular mass. Sampling below the homopause is necessary to procure a representative sample of the Venus atmosphere (with the caveat that species like \ce{H2O}, \ce{SO2}, and \ce{CO} vary with altitude even below the homopause). Different gas species have different homopause altitudes \citep{Mahieux2021}, but we expect 110 km to be a sufficiently low altitude to sample beneath the homopause for gases of interest.

At the surface of Venus, the pressure is expected to be $\approx92$~bar and the temperature is expected to be $\approx735$~K \citep{bullockgrinspoon2001}. At $\approx110$~km in altitude, where sampling is expected to occur for VATMOS-SR, the pressure and temperature are expected to be $\approx\num[round-precision=1]{1.3e-6}$~bar and 195~K~\citep{ref:mahieux2012, ref:mahieux2015, limayeetal2017, sanchezlavega2017}. The scale height at the surface is approximately 16 km, and the mean lapse rate is $\approx8$~K/km~\citep{sanchezlavega2017}. Variations in the density of the atmosphere with altitude are constrained by orbiter and ground-based telescope observations \citep{limayeetal2017}, and with general circulation models where observations are sparse or unavailable \citep{navarroetal2021,gillietal2021}. 
Approximate values consistent with observational constraints are used for the current simulations. As upcoming space missions \citep{Widemann2023}, including VERITAS \citep{Smrekar2022}, DAVINCI \citep{garvin2022} and EnVision \citep{ghail2017envision}, improve our knowledge of the structure, dynamics, and composition the Venus atmosphere, our expected freestream conditions may be updated accordingly. 

\section{DSMC Simulations}
\label{sec:dsmcsimulations}

\subsection{Modeling Assumptions}
\label{subsec:modelingassumptions}

DSMC is a numerical Monte Carlo method to solve the time-dependent nonlinear Boltzmann equation. It is a probabilistic simulation of molecular processes based on the kinetic theory of dilute gases~\citep{bird1994molecular}. The general Boltzmann equation follows the following form (with physical interpretations of the various terms included)
\begin{equation}
    \label{eq:DSMC}
    \underbrace{\frac{\partial f}{\partial t} + \frac{\vec{p}}{m} \cdot \nabla{f} + \vec{F} \cdot \frac{\partial f}{\partial \vec{p}}}_{\text{molecular motion and 
force-induced acceleration}} = \underbrace{\left(\frac{\partial f}{\partial t}\right)_{coll}}_{\text{molecular collisions}}
\end{equation}
where $f$ is a probability density function, $\vec{p}$  is the momentum vector, $\vec{F}$ is the force field acting on the particles in the fluid, and $m$ is the mass of the particle. In the DSMC algorithm, the Boltzmann equation is decoupled into two parts: molecular advection and collisions. Particles are tracked in the domain based on their position in virtual cells. Each numerical particle represents an $F_{N}$ number of real molecules. The time step is chosen such that it is smaller than the mean collision frequency. At each time step, following their advection, a number of binary pairs of particles are selected for collisions within each grid cell. Each grid cell is chosen to be smaller than the mixture mean free path.

In the collision procedure, we follow the no-time-counter (NTC) procedure of Bird~\citep{bird1994molecular} for the selection of collision pairs. To reduce statistical scatter and ensure that collisions are performed accurately, it is generally recommended that a minimum of 10 to 20 particles per computational cell are used. SPARTA uses the Variable Soft Sphere (VSS) model to simulate particle interactions~\citep{koura1991variable}. High-temperature fits to the collision integrals (CI) for transport properties computations for Venusian atmosphere~\citep{ref:wright2007, swaminathan2015consistent} used in this study are provided in Appendix~\ref{sec:collisionmodelparameters}. Additionally, the Mars/Venus set of chemical reactions developed by Johnston and Brandis~\citep{ref:johnston2014} is used in the Total Collision Energy (TCE) model of Bird~\citep{bird1994molecular}. Some Arrhenius rates from the Johnston and Brandis model had to be refitted such that the Arrhenius temperature exponent fits in a certain range, as required by the TCE model (described in~\citep[Ch. 7]{higdon2018monte}). The full set of Arrhenius rates can be found in Appendix~\ref{sec:chemical_rates}.

\subsection{Freestream Conditions}
\label{subsec:freestreamconditions}

As mentioned in Sec.~\ref{sec:venusatmosphere}, Venus' atmosphere is approximately composed of 96.5\,\% \ce{CO2}, 3.5\,\% \ce{N2}, and trace amounts of other gases. Depending on the noble gas of interest, the concentrations are expected to be between a few ppbv and $\approx150$~ppmv, which corresponds to mole fractions between $\approx10^{-4}$ and $\approx10^{-9}$. Given these extremely low mole fractions, all the noble gases are considered to be trace species compared with the reactive species comprising the Venusian atmosphere. Simulating the expected mole fractions of the noble gases with an acceptable accuracy in our DSMC model would prove computationally impossible as it would require using $\approx10^{17}$ numerical particles. Therefore, in all the Cupid's Arrow and VATMOS-SR numerical studies, the mole fractions of the noble gases were artificially increased. To account for possible reactions between the reactive freestream gases, additional species (\ce{CO}, \ce{O}, \ce{N2}, \ce{NO}, \ce{N}, \ce{C}, \ce{CN} and \ce{C2}) are also considered in the mixture of interest. Simulations including ionized species (CO+, O+, N+, NO+, O2+, C+, ionized noble gases, and e-) at 10.5~km/s entry conditions were performed. Due to the large ionization potentials of noble gases, no meaningful levels of ionization were observed for them. For non-noble gas species, only CO+, NO+ and C+ showed any non-trivial levels of ionization. However, the total mole fractions of ions and electrons were an order of magnitude lower than 1\%. We therefore concluded that, under these conditions, including ionization would not provide any additional physics, but would provide a significant computational slowdown. However, should faster entry conditions be investigated (e.g., 13 km/s), we acknowledge that ionization and possibly electronic excitation could become an important phenomenon that should be accounted for in future studies. Based on these results, a neutral gas mixture is assumed in the following sections.

In particular, elemental fractionation refers to the enrichment of one element compared to another, and isotopic fractionation refers to the enrichment of one isotope of a species compared to another isotope of the same species, all with respect to the naturally occurring composition in the Venus atmosphere. The primary contributor to gas fractionation is differential diffusion, which is driven by the differences in molecular mass of the species present in the gas mixture and is expected to mainly occur in the shock and compression layers in front of the vehicle. The gas expansion, as the mixture travels through the valve orifice and into the sampling tank, is also postulated as a possible secondary fractionation contributor. Sensitivity studies were conducted in~\citep{ref:rabinovitch2019a} to quantify the impact of the assumed noble gas mole fraction in the freestream of the DSMC simulations on their fractionation, particularly focusing on the standard deviation of mixture mole fraction in the sampling tank. The preliminary conclusion was that freestream mole fractions for all noble gases were required to be a minimum of 0.1\,\%, with a value of 1\,\% preferred. Further studies (outside the scope of this paper) showed that a 1\,\% freestream mole fraction for the noble gas species provided the best compromise between accuracy for the concentration of the trace noble gas species in the sample acquisition tanks, while also minimizing the effect the increased noble gas species have on changing the Venusian atmospheric transport properties. Furthermore, we verified that simulations that modeled each noble gas species separately recovered the same fractionation value as simulations that modeled the various species together (within the random fluctuations). This independent treatment of noble gases can also be justified by the fact that the actual noble gas concentrations in the Venusian atmosphere are extremely low which implies that the probability of inter-noble gas collisions is negligible.

The determination of the noble gas isotopic ratios is a primary scientific objective, and the required precision for these measurements is shown in  Table~\ref{fig:science-goals}. To investigate the mass-dependent fractionation trends observed with the simulations, a transfer function between the ratios in the tank and those in the freestream is also generated. For all the various isotopes of each noble gas, identical mole fractions are specified in each of the fifteen independent simulations. Therefore, comparing the number densities of the various isotopes in the tank will provide a direct measure of the fractionation. We also confirmed that the tank pressure was similar for all instances in the fifteen independent simulations (within 0.6\,\%).

Table~\ref{tab:sim-matrix} summarizes the baseline simulation matrix used for this study.
The various isotopes that are considered are: $^{128}$Xe, $^{129}$Xe, $^{130}$Xe, $^{132}$Xe, $^{136}$Xe, $^{3}$He, $^{4}$He, $^{36}$Ar, $^{38}$Ar, $^{40}$Ar, $^{20}$Ne, $^{22}$Ne, $^{82}$Kr, $^{84}$Kr and $^{86}$Kr.
All isotopes of a specific noble gas species are modeled with the same transport properties (i.e., VSS parameters) except for their masses. While there is no available data regarding the transport properties of various isotopes of a noble gas at high temperature, since different isotopes of the same gas have the same electron cloud, the interatomic potentials are presumed to be identical, and therefore the  VSS parameters should be the same for all isotopes of a noble gas. Therefore, only the mass of isotopes should affect their transport properties.
The value of the freestream temperature is 195~K. The baseline freestream velocity and number density used in this study are 10.5~km/s and \num[round-precision=2]{4.84e19}~m$^{-3}$, respectively. At those conditions, the corresponding Mach number is 49.5. Furthermore, the effects of varying the freestream density (increasing and decreasing it by 25\,\% when compared to the baseline value) as well as varying the freestream velocity (a second velocity of 13.0~km/s, which corresponds to the velocity the spacecraft would have if it were a secondary payload to the Dragonfly~\citep{lorenz2018dragonfly} trajectory, that included a Venus flyby) are investigated. 

\begin{table}
    \centering
    \caption{Baseline simulation matrix (15 independent simulations) used for this study. Only the mole fraction of the noble gas species is varied between each simulation.}\vspace{1mm}
    \resizebox{0.9\columnwidth}{!}{\begin{tabular}{||c||c|c|c|c|c|c||}
            \hline\hline
            Case & U$_\infty$ (km/s)                                                                                                & T$_\infty$ (K) & n (m$^{-3}$) & X$_{CO_{2}}$ & X$_{N_{2}}$ & X$_{i}$ \\
            \hline
            i    & 10.5                                                                                                             & 195            & \num{4.84e19}      & 0.937        & 0.053       & 0.01    \\
            \hline
            i=1- & \multicolumn{6}{c||}{X$_{i}$=\{$^{128}$Xe, $^{129}$Xe, $^{130}$Xe, $^{132}$Xe, $^{136}$Xe, $^{3}$He, $^{4}$He, }                                                                        \\
            15   & \multicolumn{6}{c||}{$^{36}$Ar, $^{38}$Ar, $^{40}$Ar, $^{20}$Ne, $^{22}$Ne, $^{82}$Kr, $^{84}$Kr, $^{86}$Kr \} }                                                                        \\
            \hline\hline
        \end{tabular}}
    \label{tab:sim-matrix}
\end{table}

\subsection{Simulations Parameters}
\label{subsec:simulationsparameters}

Particle--surface collisions are modeled using a diffuse reflection model. Two spacecraft surface temperature conditions are investigated in this study: a cold wall surface temperature fixed at 295~K, which was based on the expected temperature inside the probe in proximity to Venus, and a second more complex, variable surface temperature model, which makes use of the radiative equilibrium law, and will be referred to as \say{hot wall} model. In the latter, the surface temperature is determined from the Stefan-Boltzmann law, $Q_{w} = \sigma \epsilon T_{w}^{4}$, where $Q_{w}$ represents the net heat flux on the surface, $T_{w}$ the surface temperature, $\sigma$ the Stefan-Boltzmann constant, and $\epsilon$ the emissivity (0.85 for the thermal protection system material considered in this study). No surface reactions are included in this study, and all simulation results presented in Sec.~\ref{sec:results} initialize the simulations with \num{90000} particles per cell in the freestream.

SPARTA allows the grid to be adapted either between runs or on-the-fly during a run as flow parameters or particle densities change spatially~\citep{plimpton2019direct}. Individual child grid cells can be refined to the next level in the hierarchy (\emph{e.g.}, 1 cell becomes $2 \times 2 \times 2 = 8$ smaller grid cells). The criterion for refining a grid cell is usually chosen as the mean free path between molecular collisions, which is a function of the number density and the temperature of the particles in the cell.
The maximum level of grid refinement is capped to 10, in order to keep the amount of computer memory manageable. A study on the effects of the maximum grid refinement level is provided in Appendix~\ref{sec:AMR}. In turn, following grid adaptation, a total of up to 45 million grid cells and 317 million particles are used in the simulations. In the area of the highest number density (sampling inlet, upstream of the Mindrum valve), due to the adaptive mesh refinement, this leads to around 2 particles per cell, which is below the desired number referenced previously. However, this is necessary to ensure the computational cost remains reasonable. This allows for the mean free path to be smaller than half of the cell size in a large part of the domain, except in the sampling inlet upstream of the valve. The time step is chosen as \num[round-precision=2]{2.42e-8}~s, in order to ensure it is always less than half of the mean collision time anywhere in the domain.

\begin{figure}
    \centering
    \includegraphics[width=0.4\textwidth]{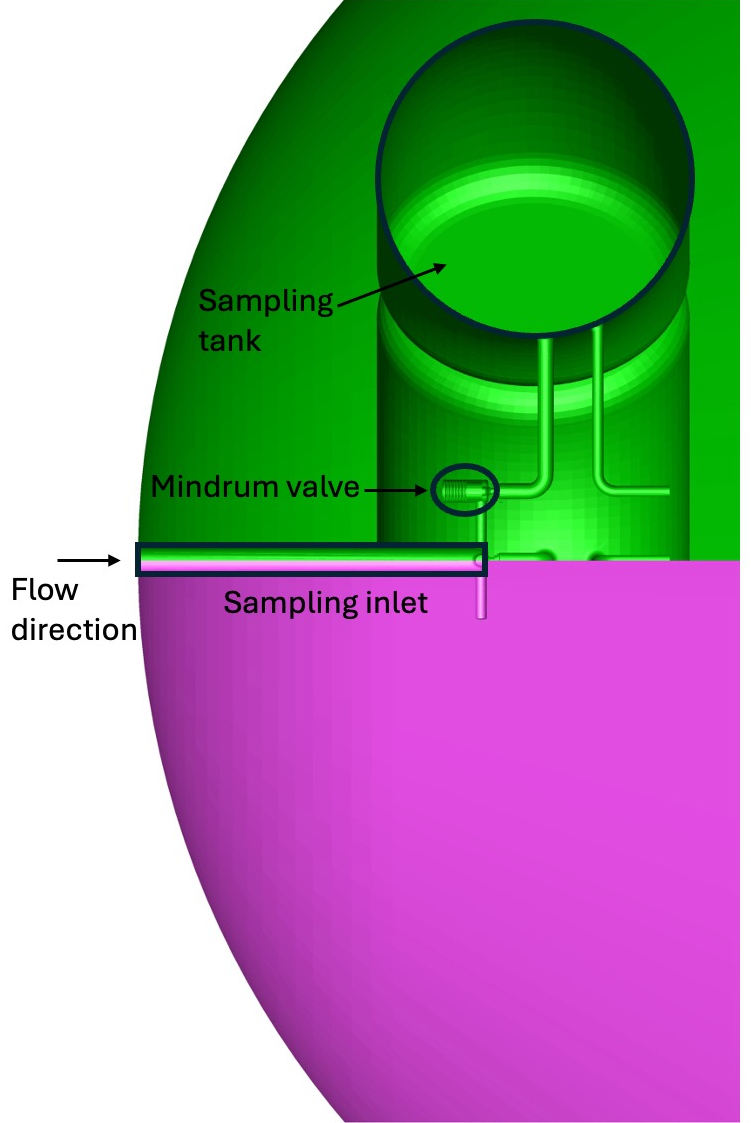}
    \caption{Zoomed-in schematic view of the (top) \say{open valve} geometry, that includes the sampling tank, and (bottom) \say{closed valve} geometry. Only a quarter of the actual geometry is simulated.}
    \label{fig:CA-surf-both}
\end{figure}

\begin{figure*}
	\begin{subfigure}{.59\textwidth} 
        \centering
        \includegraphics[width=\linewidth]{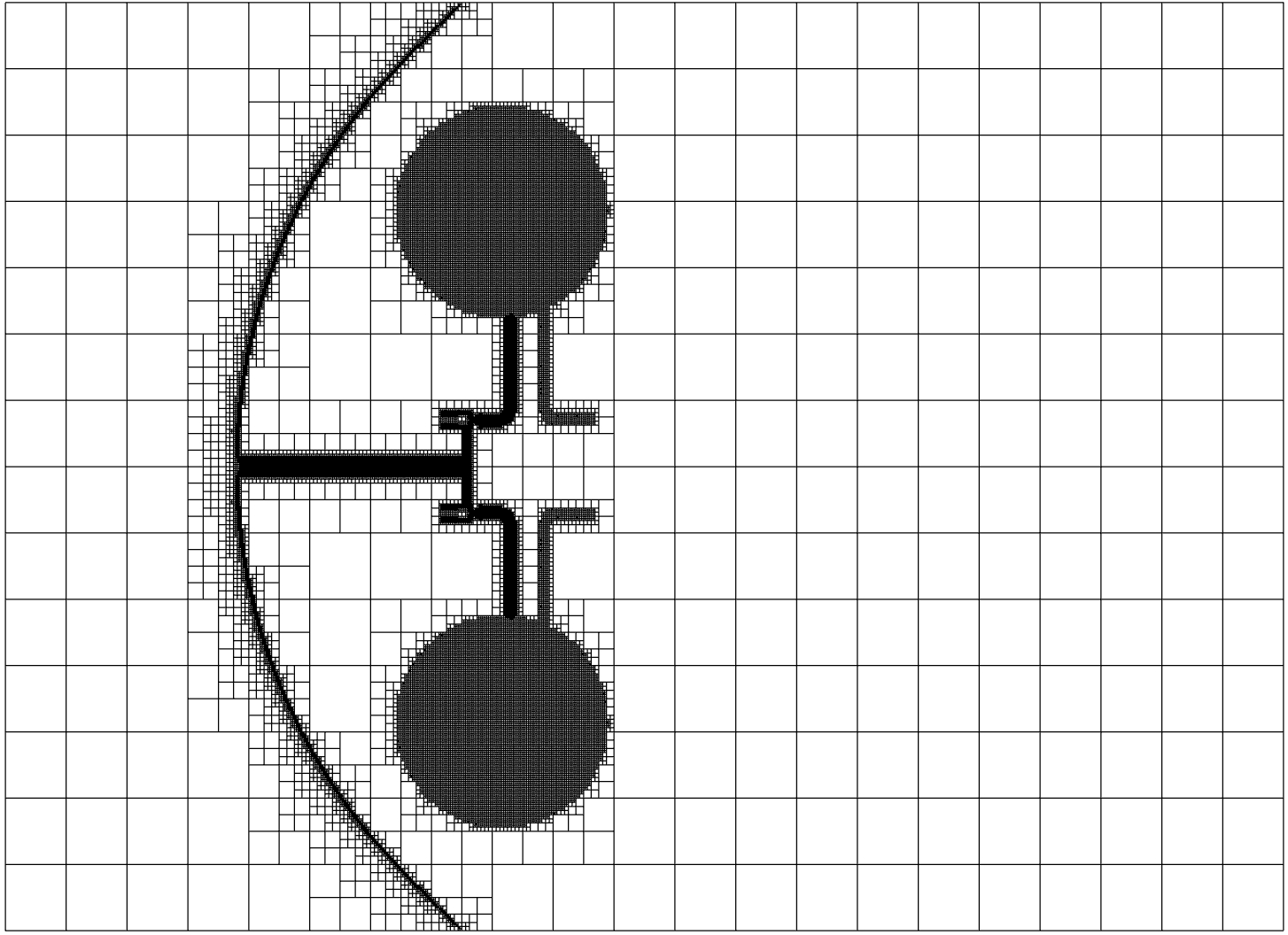}
        \caption{Entire computational domain.} \label{fig:grid-CA}
	\end{subfigure}
    \hfill
	\begin{subfigure}{.4\textwidth}
		\centering
        \includegraphics[width=\linewidth, angle=-90]{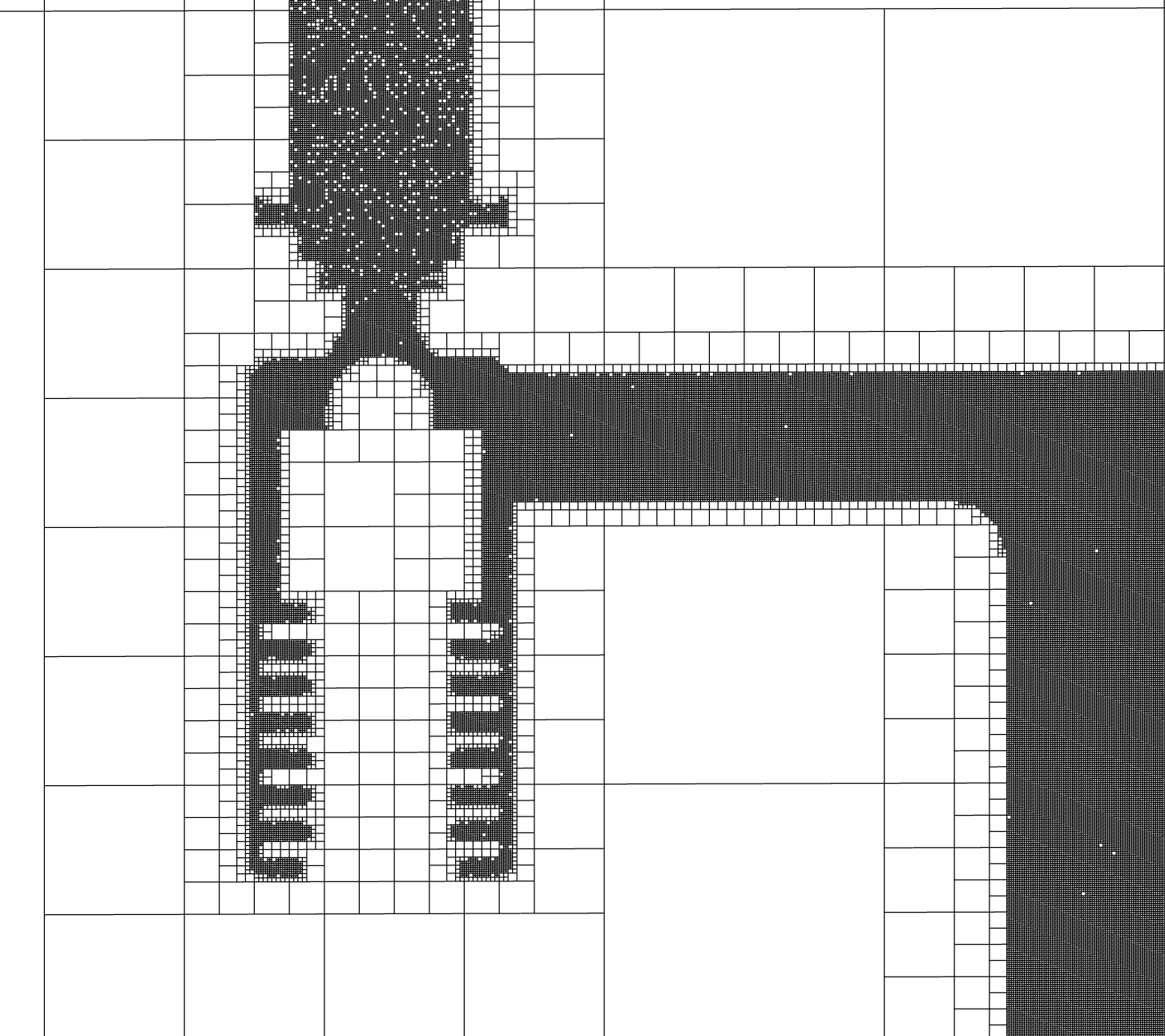}
		\caption{Close-up view on the Mindrum valve.}  \label{fig:grid-CA-zoom}
	\end{subfigure}
 		\caption{Computational grid once the simulations have reached steady state.}  \label{fig:comp-domain}
 \end{figure*}

Based on these variations, we performed a total of 126 simulations to complete the full simulation matrix. These simulations are computationally expensive since they need to cover many orders of magnitude for the number density between the freestream and the valve area. To reduce the computational cost, a quarter of the actual VATMOS-SR probe is simulated, with symmetry boundary conditions applied in the two directions perpendicular to the flow direction. For each of the simulations, 1280 AMD EPYC Rome cores (10 nodes with 128 cores/node) are used on the NASA Supercomputer Aitken. The total available memory is 5.12~TB, and all simulations are run for 120 hours, which is the maximum allowable wall-time.
For each simulation, the Mindrum valve is first placed in a closed state, and steady-state is allowed to establish in front of the spacecraft and inside the tube ahead of the valve. That initial run takes place for 1.5 million time steps or 36.39~ms. We confirmed that the number of particles as well as collision cells have reached a steady state after the first one million time steps by studying the evolution of the number of particles, particle-particle and particle--surfaces collisions, and the number of collision cells, in the domain. The valve is then numerically opened by replacing the entire surface with a new surface geometry containing the open valve and sampling tank. The flow is then allowed to establish inside the valve, secondary tube, and sampling tank. The simulation is then continued for approximately another 5 million time steps or 120~ms.
Fig.~\ref{fig:CA-surf-both} shows the initial \say{closed valve} geometry, on top in green, as well as the second \say{open valve} geometry, at the bottom in cyan. The sampling inlet, Mindrum valve, as well as the sampling tank are all identified.

Number densities for each species present in the sampling tank  are output at a frequency of \num{100000} time steps, or every 2.42~ms.
Fig.~\ref{fig:grid-CA} shows an excerpt of the computational grid on a quarter of the geometry following adaptive refinement. One can visually observe the large number of computational cells needed in regions of high density (compression layer, pipes, and valves). Fig.~\ref{fig:grid-CA-zoom} shows a close-up view of the grid in the Mindrum valve region, showing how the complex internal geometry of the valve is captured in these simulations.

Although previous studies~\citep{ref:rabinovitch2019a} focused on 2-dimensional axisymmetric simulations, all simulations for the VATMOS-SR spacecraft in this study are performed on a 3-dimensional geometry. While 2-dimensional simulations have a computational cost that is an order magnitude less than the 3-dimensional simulations, and are entirely capable of modeling the fractionation through the bow shock, compression layer, and sampling inlet, the inability to simplify the complex flow path through the Mindrum valve and various turns in an axisymmetric manner, means that the 2-dimensional simulations are not entirely capable of predicting the same fractionation as the 3-dimensional ones.
\section{Results and Sensitivity Studies}
\label{sec:results}
In this section, we focus on the results of the 3-dimensional parametric studies investigating the influence of various parameters on the fractionation of noble gases for VATMOS-SR. In Appendix~\ref{sec:fractionation}, we provide some background on the fractionation of noble gases as well as its modeling with DSMC. In particular, we perform some normal shock simulations and investigate the fractionation of noble gases in these shocks. As opposed to the normal shock, the simulations of the blunt-body flow incorporate pressure diffusion effects in addition to thermal diffusion. For the results presented in this section, we model the full spacecraft geometry, including the sampling system. While the primary quantity of interest is the mixture composition in the sampling tank, we also provide an overview of the flow field. We start by analyzing the bow shock and compression layer, which are crucial in fractionating the noble gases. We then study both the elemental and isotopic fractionation of the noble gases of interest for the baseline conditions, which are the ones we expect the spacecraft to encounter during a nominal Venus atmospheric pass. Subsequently, we take a look at the effects of statistical scatter, since the DSMC represents a large amount of real molecules by a small number of simulated molecules, and therefore requires time averaging. Finally, we investigate the effects of the surface boundary conditions in the numerical model as well as the freestream number density and velocity.

\subsection{Flow Field and Stagnation Line}
\label{sec:flowfield}

\begin{figure*}
	\begin{subfigure}{.49\textwidth} 
        \centering
        \includegraphics[width=\linewidth]{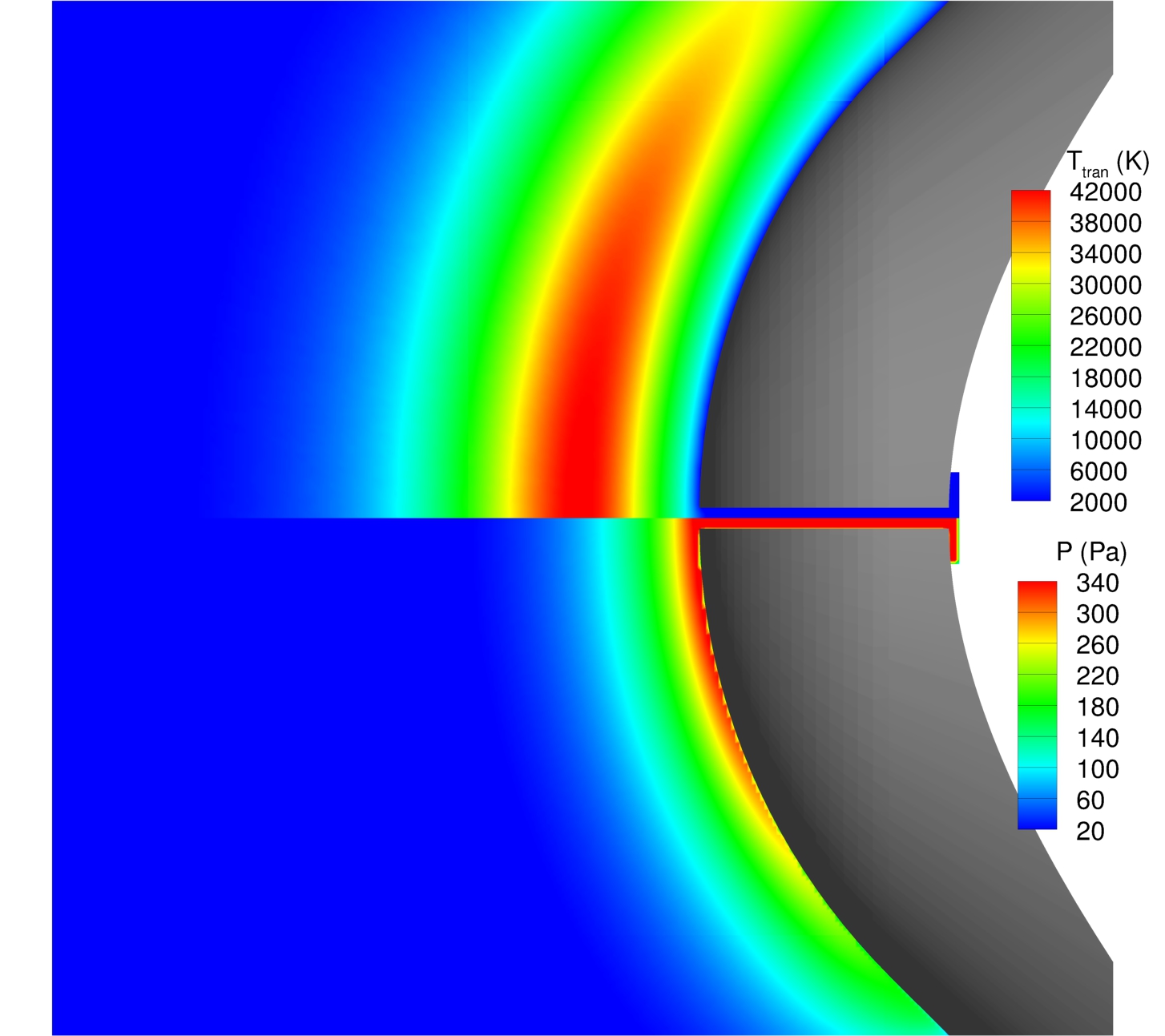}
        \caption{Contours.} \label{fig:contour-T-P}
	\end{subfigure}
    \hfill
	\begin{subfigure}{.49\textwidth}
		\centering
        \includegraphics[width=\linewidth]{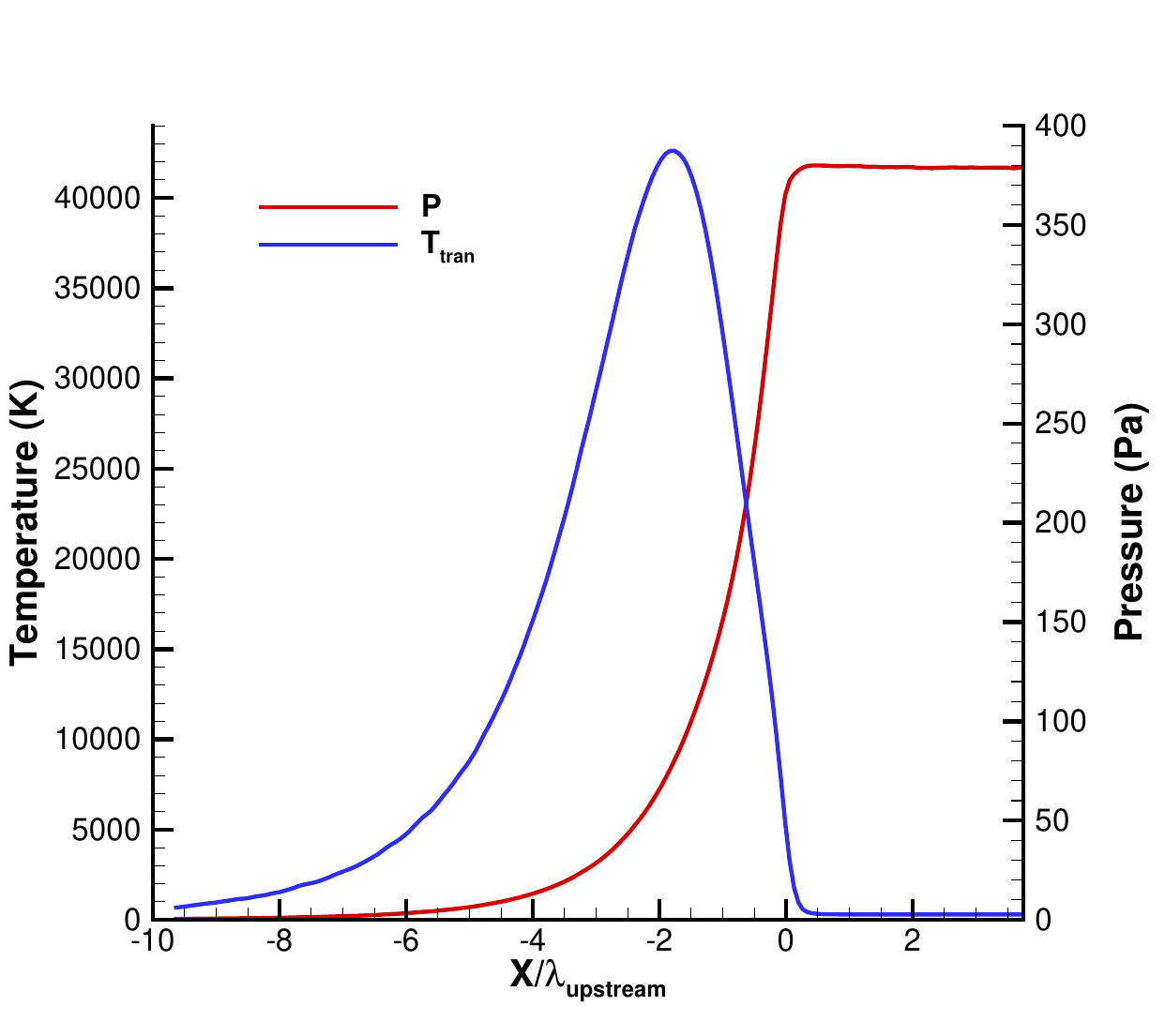}
		\caption{Stagnation line plot.}  \label{fig:line-plot-T-P}
	\end{subfigure}
 		\caption{Translational temperature and pressure for the baseline conditions. The view is zoomed-in on the nose of VATMOS-SR vehicle, where the sampling occurs, and the bow shock and compression layer region, as well as the sampling inlet.}  \label{fig:3D-baseline-flow}
 \end{figure*}

We perform simulations of the VATMOS-SR spacecraft at the baseline conditions shown in Table~\ref{tab:sim-matrix}. We first focus on the \say{closed valve} results, \emph{i.e.}, up to the end of the sampling inlet. The results in this section use a constant, cold wall surface temperature of 295~K. Contour and stagnation line plots for the translational temperature and pressure are shown in Fig.~\ref{fig:3D-baseline-flow}. The contour plot shows the bow shock is well developed around the blunt body, and that the peak temperature is located $\approx2$ mean free path from the surface. The peak translational temperature reaches \num{42700}~K. Furthermore, the stagnation line plot shows that temperature drops rapidly inside the inlet to under 350~K in a distance of only one mean free path, or 25~mm.
Additionally, the pressure reaches a maximum value of $\approx380$~Pa near the entrance of the inlet, then plateaus inside the inlet. This pressure value is slightly higher, as expected (since the flow velocity decreases to nearly zero near the inlet), than the post-shock pressure of 341~Pa for the DSMC simulations of a normal shock described in Appendix~\ref{sec:fractionation}.

In Fig.~\ref{fig:3D-baseline-mass-frac}, we focus on the lightest and heaviest noble gases, $^{3}$\ce{He} and $^{136}$\ce{Xe}, which are each modeled in an independent simulation. 
One can see that the two gases exhibit opposite trends after starting at a 1.0 normalized mass fraction in the freestream. $^{3}$\ce{He} increases along the stagnation streamline, reaches a peak of approximately 1.42 in the shock, and then the value drops to a minimum around 0.15 at the start of the sampling inlet, before slowly rising to 0.26.
$^{136}$\ce{Xe} decreases along the stagnation streamline, reaches a minimum around 0.65 in the shock, and then the value sharply increases in the compression layer, reaching a peak around 5.77 at the start of the sampling inlet, before dropping to 1.67 further downstream.
Those conclusions are similar to those of Bird~\citep{bird1994molecular}, as summarized in Appendix~\ref{sec:fractionation}, and show that light and heavy noble gas (with light and heavy defined with respect to the mixture's mean molecular weight) behave in opposite ways, and while the initial effect seen along the stagnation line is due to thermal diffusion in the shock, the pressure diffusion taking place in the compression layer has a stronger effect. Therefore, for a mixture of noble gases that have a low concentration compared to the reactive \ce{CO2}-\ce{N2} mixture, one expects that the hypersonic flow around a blunt body will concentrate the heavier gases and deplete the lighter ones at the surface and in the sampling inlet.

 \begin{figure*}
	\begin{subfigure}{.49\textwidth} 
        \centering
        \includegraphics[width=\linewidth]{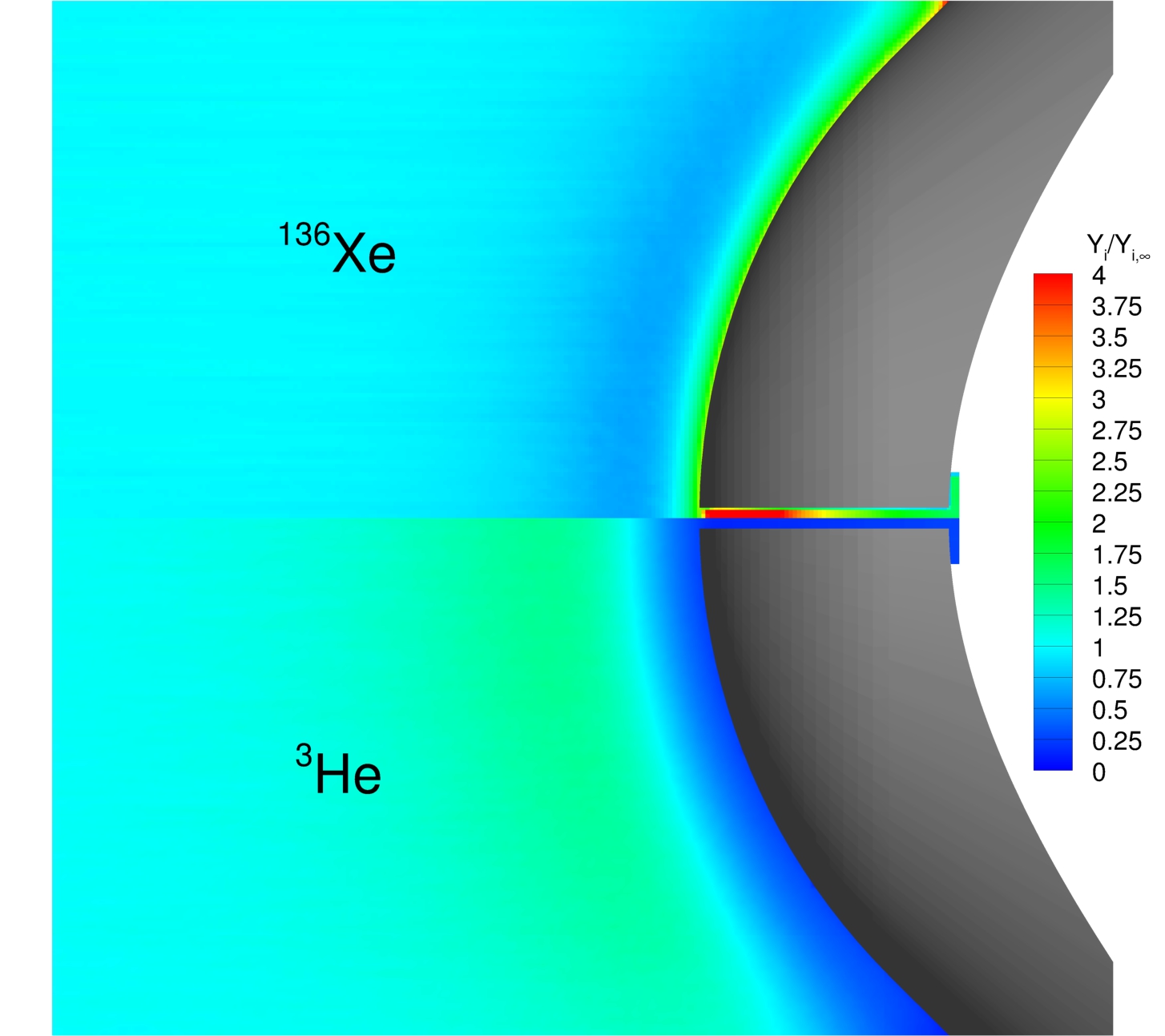}
        \caption{Contours.} \label{fig:3D-comp-contour}
	\end{subfigure}
    \hfill
	\begin{subfigure}{.49\textwidth}
		\centering
        \includegraphics[width=\linewidth]{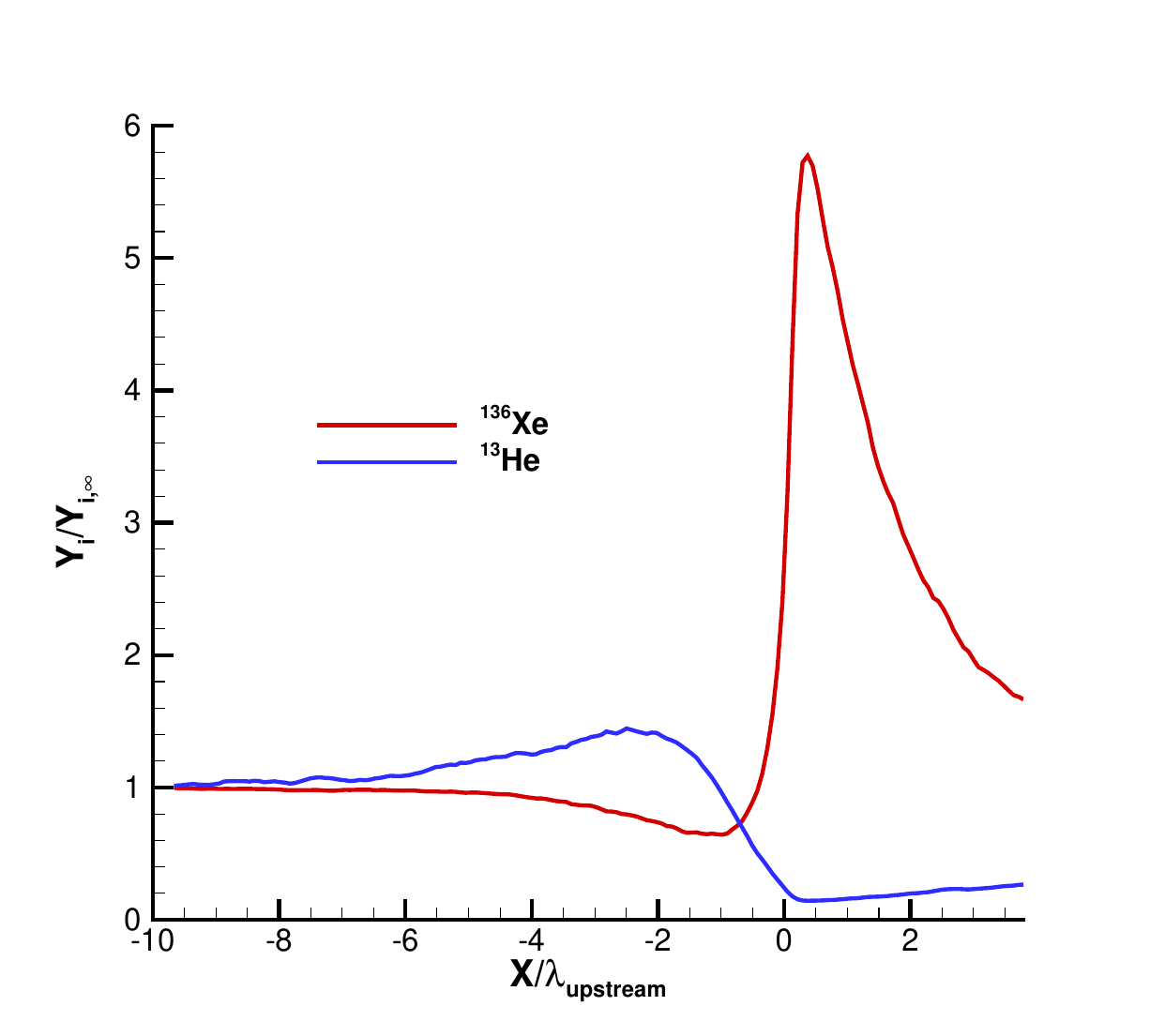}
		\caption{Stagnation line plot.}  \label{fig:3D-comp-line-plot}
	\end{subfigure}
 		\caption{Normalized mass fractions of $^{136}$\ce{Xe} and $^{3}$\ce{He} for the baseline conditions. The view is zoomed-in on the nose of VATMOS-SR vehicle, where the sampling occurs, and the bow shock and compression layer region, as well as the sampling inlet.}  \label{fig:3D-baseline-mass-frac}
 \end{figure*}

\subsection{Sampling Tank}
\label{sec:tank}
The goal of this study is to demonstrate that it is possible to relate the relative concentrations of noble gas species in the tank to that in the freestream using DSMC simulations. It is known that the design of a future sampling system may differ from the one assumed here; however, incorporating the major components of a sampling system -- particularly the Mindrum valve -- into our geometry demonstrates that fractionation can be predicted for realistic flight-like geometries and hardware. This strengthens our confidence that DSMC simulations can be applied to future sampling systems to quantify fractionation as the spacecraft design evolves. Due to the temperatures encountered in the shock, dissociation of \ce{CO2} and \ce{CO} are the predominant chemical reactions taking place in front of the vehicle. For the baseline conditions, the mixture in the sampling tank is composed of \ce{CO}~$\approx56$\,\%, \ce{O}~$\approx23$\,\% and \ce{O2}~$\approx12$\,\%, by mole. The remaining fraction of species in the tank consists of \ce{CO}, \ce{N2}, \ce{N}, \ce{NO}, and the noble gas species of interest. It is important to note that despite no \ce{C}, \ce{C2}, and \ce{CN} being recorded in the sampling tank, those species do serve as intermediates in the chemical reactions occurring in the shock and compression layer, and are therefore important to include in the chemical mechanism.

The pressure upstream of the Mindrum valve at the end of the closed valve simulation is $\approx380$~Pa, and the sample tank is assumed to be at vacuum once the open valve simulation starts. At the end of the open valve DSMC simulations, the sample tank has reached a pressure of $\approx5$~Pa. While the pressure in the tank would continue to increase if the simulations were run longer, the simulations are stopped at this point due to computational cost. A linear change in pressure with respect to time in the tank is observed in this regime as expected, due to the large pressure ratio between the upstream and downstream side of the Mindrum valve (the orifice inside the valve acts as a chokepoint in the flow). VATMOS-SR is expected to operate in a regime where the flow remains choked at the valve orifice.

\begin{figure}
    \centering
    \includegraphics[width=0.6\columnwidth]{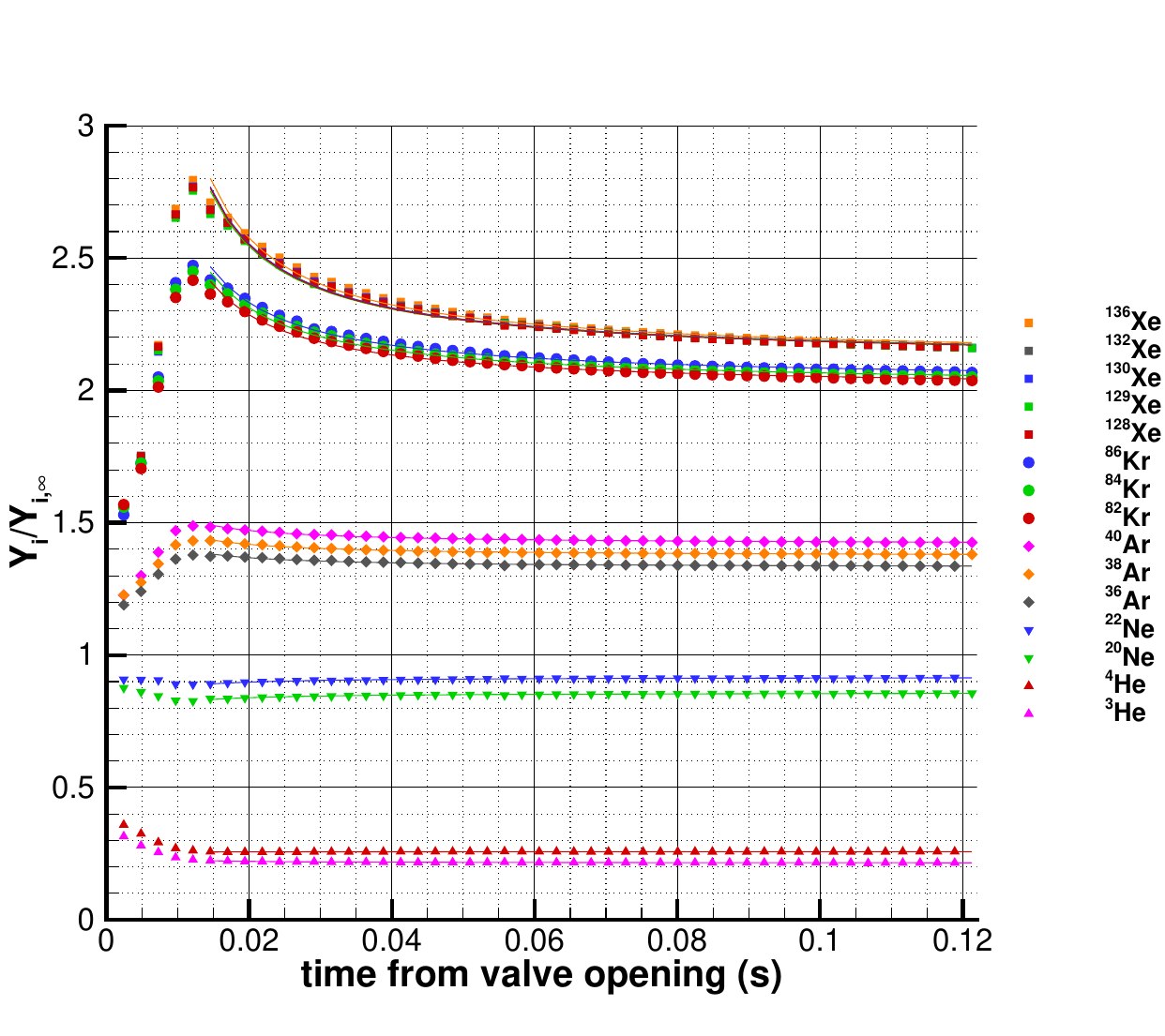}
    \caption{Normalized mass fractions of the fifteen noble gas species in the sample acquisition tank as a function of time from valve opening for the baseline conditions. Solid lines show the Michaelis-Menten fit used to extract the steady-state value of the normalized mass fractions.}
    \label{fig:mass-fracs-vs-time}
\end{figure}

The normalized mass fraction of noble species $i$, $Y_{i}/Y_{i,\infty}$, is defined as the mass fraction of a noble species normalized by the mass fraction of the same noble species in the freestream, and is computed as:
\begin{equation}
    \frac{Y_{i}}{Y_{i,\infty}} = \frac{M_{i} \cdot n_{i,tank}} {\sum\limits_{j=1}^{N_{species}} M_{j} \cdot n_{j,tank}} \frac{\sum\limits_{j=1}^{N_{species}} M_{j} \cdot n_{j,\infty}} {M_{i} \cdot n_{i,\infty}},
    \label{eq:ele-frac}
\end{equation}
where $M_{j}$ and $n_{j}$ are the molecular weight and number density of species $j$, respectively. The $tank$ and $\infty$ subscripts represent values in the sampling tank and freestream, respectively.
The normalized mass fractions of each noble gas species are plotted as a function of time in Fig.~\ref{fig:mass-fracs-vs-time}. It can be seen that after an initial transient time on the order of 40~ms for the heavier species, and 10~ms for the lighter ones, the normalized mass fraction ratios asymptote to a steady-state value. This steady-state value is then used as the fractionation ratio for each species, and we use a Michaelis-Menten~\citep{michaelis1913gerak,johnson2011original} type hyperbolic fit to extract the exact steady-state value for each species. Extrapolating to a steady state value saves computational cost (by enabling a shorter overall simulation duration), and it is assumed that a spacecraft would sample for order(s) of magnitude more time than what can be simulated with a reasonable computational cost, justifying the steady-state fitting in this work.

\begin{figure}
    \centering
    \includegraphics[width=0.6\columnwidth]{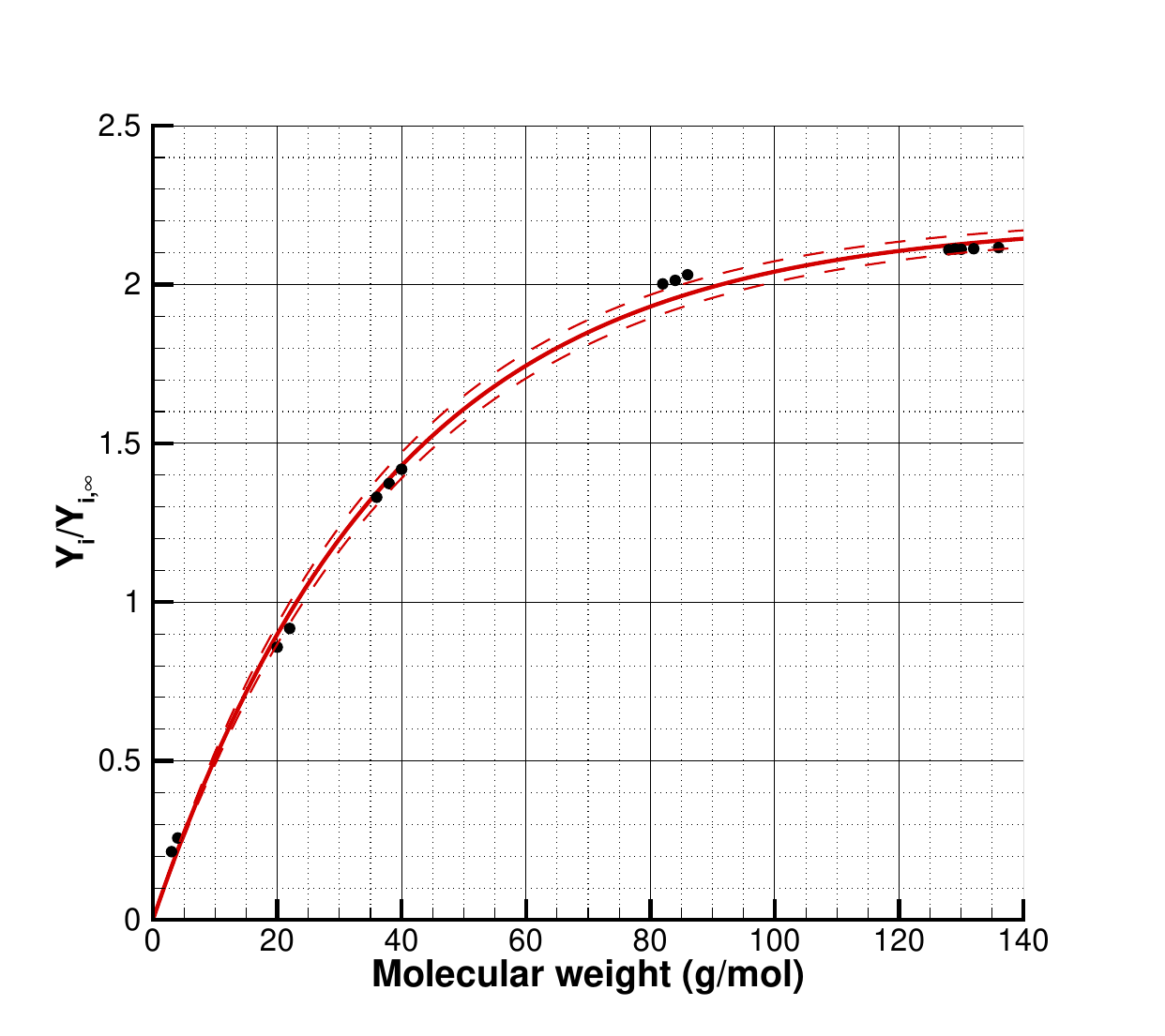}
    \caption{Elemental fraction as a function of molecular weight of the noble gas for the baseline conditions. Also shown is a fit to the data using a one-phase exponential association equation. The dashed red lines represent the boundaries of the fit due to the uncertainties on the 2 fitting parameters.}
    \label{fig:elemental-fractionation}
\end{figure}

Fig.~\ref{fig:elemental-fractionation} shows the elemental fractionation steady-state values, derived from Fig.~\ref{fig:mass-fracs-vs-time}, fitted with the one-phase exponential association equation:
\begin{equation}
    \frac{Y_{i}}{Y_{i,\infty}} = \frac{Y_{i}}{Y_{i,\infty, max}} \left( 1 - \exp \left(\frac{-M}{M_{0}} \right) \right),
    \label{eq:ele-frac-fit}
\end{equation}
where $M$ is the molecular weight of the noble gas, $M_{0}$ is the molecular weight scale parameter and $Y_{i}/Y_{i,\infty, max}$ is the asymptote. The latter is defined by the sampling scenario specifics, such as atmospheric composition, bow shock conditions, species dissociation, species transport through hardware, etc. The term in the parentheses can be interpreted as a cumulative distribution function in analogy with the standard Weibull distribution for the special case of the shape parameter equal to one and a scale parameter $M_{0}$. A shape parameter of one, $\left(-M/M_{0}\right) ^1$, attests to the randomness of the process, inferring constant fractionation probability per unit mass. The exponential term, $\exp\left(-M/M_{0}\right)$, can take on a meaning analogous to the Weibull reliability function, describing the transport behavior as a function of the species molecular weight. The least-square fit to the data yielded $M_{0}=38.1 \pm 1.3$~g/mol, and $Y_{i}/Y_{i,\infty, max}=2.20\pm0.02$. The fit function crosses $Y_{i}/Y_{i,\infty, max}=1$ at $M=23.1 \pm 1.0$~g/mol (henceforth referred to as the \say{tipping point}), which is in a reasonably good agreement with the average molecular weight of the mixture in the tank, $M_{t}=26.63 \pm 0.73$~g/mol (the uncertainty on the latter number is calculated as the difference between the average molecular weight of the mixture in the tank for mixtures containing the heaviest noble gas, $^{136}$\ce{Xe}, and the lightest noble gas, $^{3}$\ce{He}). Of note, is that the error bars on the symbols, computed from the slope of the Michaelis-Menten fit, are smaller than the size of the symbols, and therefore not depicted.

\begin{figure*}
	\begin{subfigure}{.32\textwidth} 
        \centering
        \includegraphics[width=\linewidth]{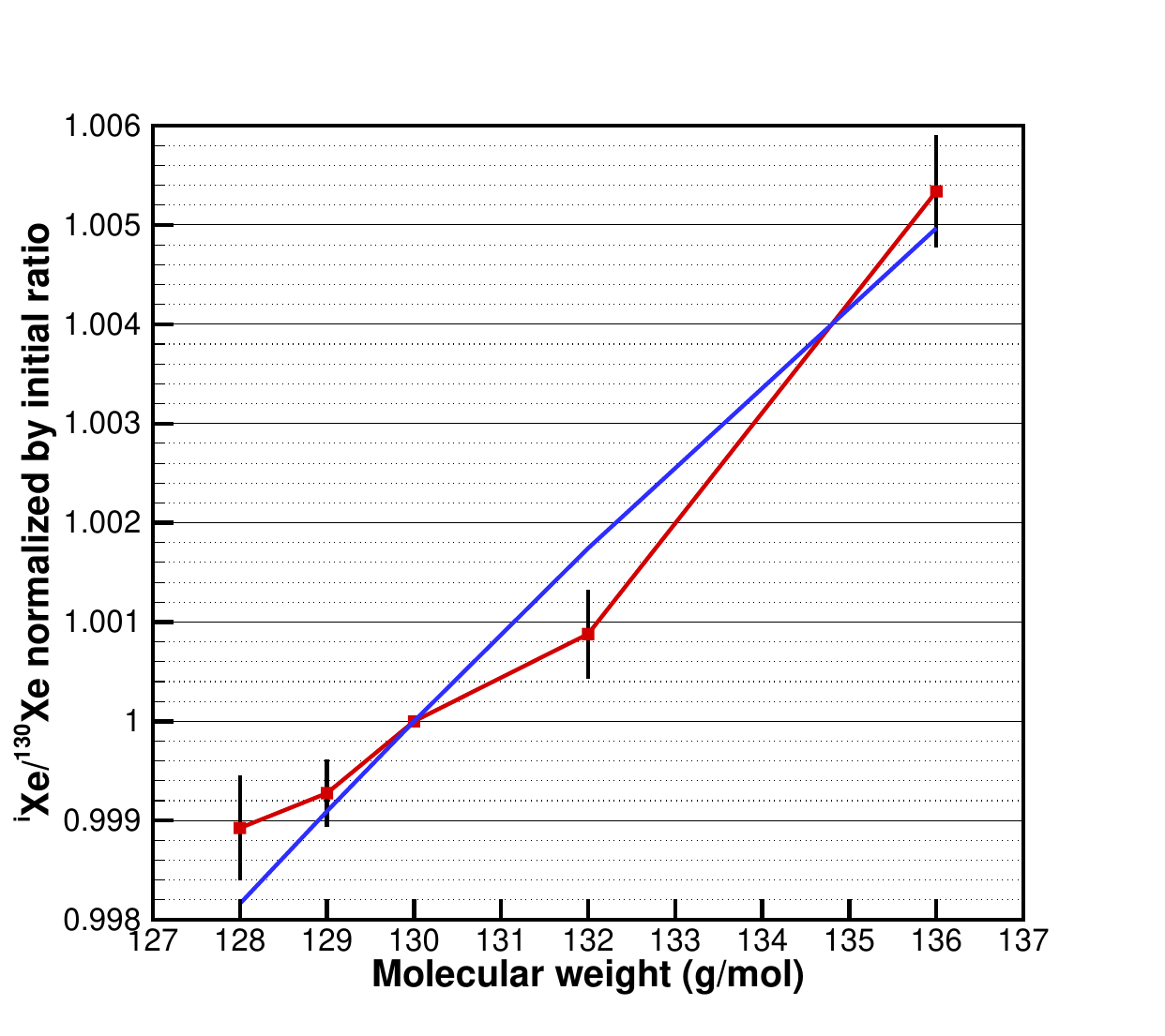}
        \caption{\ce{Xe}} \label{fig:Xe130-ratios}
	\end{subfigure}
    \hfill
	\begin{subfigure}{.32\textwidth}
		\centering
        \includegraphics[width=\linewidth]{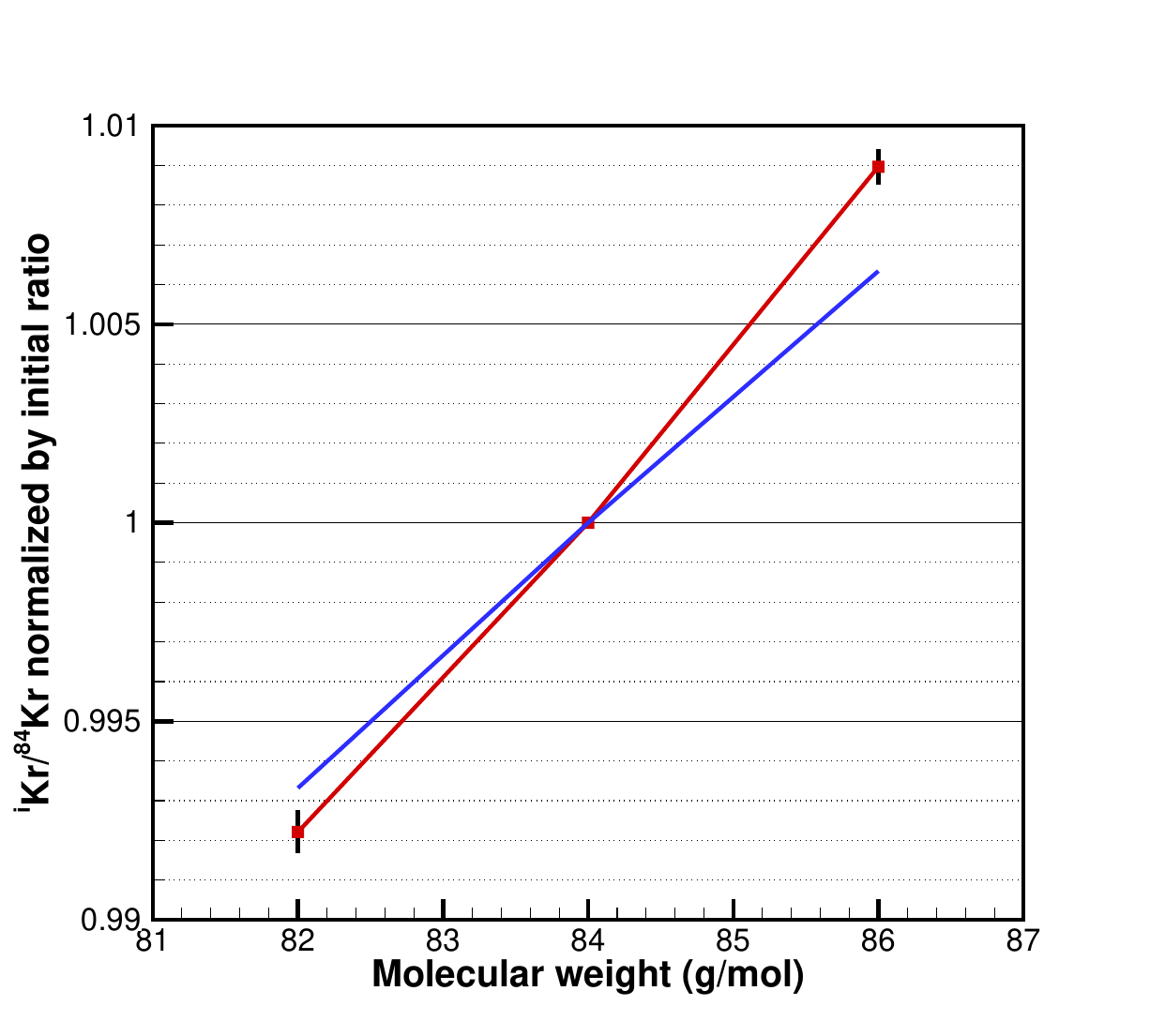}
		\caption{\ce{Kr}}  \label{fig:Kr84-ratios}
	\end{subfigure}
    \hfill
	\begin{subfigure}{.32\textwidth}
		\centering
        \includegraphics[width=\linewidth]{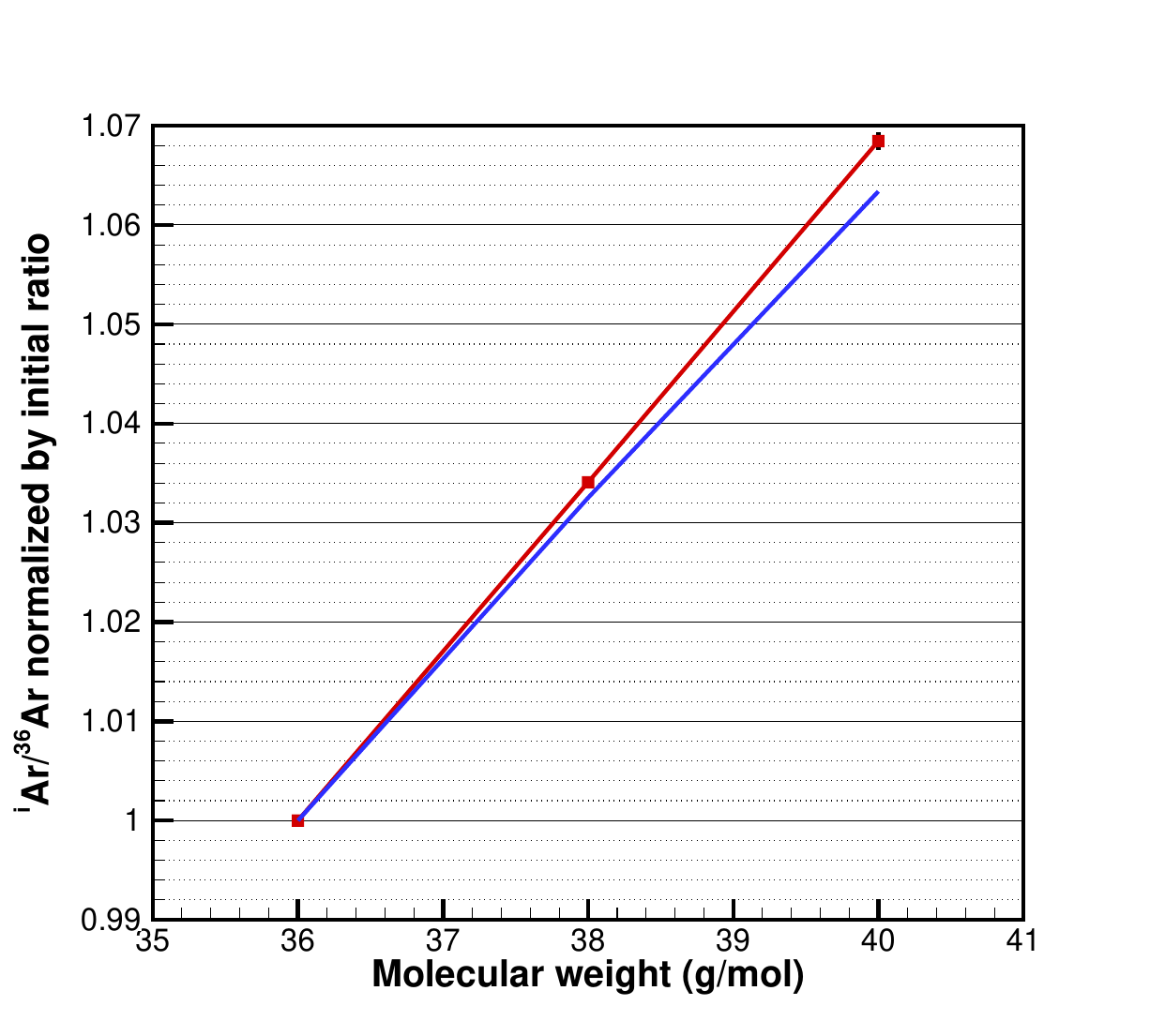}
		\caption{\ce{Ar}}  \label{fig:Ar36-ratios}
	\end{subfigure}
 		\caption{The red line and symbols represents the isotopic, molar fractionation for various isotopes of one noble gas, when compared to a reference isotope. The black error bars represent one standard deviation from the average. The blue lines represent the fit from eqn.~\ref{eq:ele-frac-fit}.}  \label{fig:isotopic-ratios-plot}
 \end{figure*}

One major conclusion from Fig.~\ref{fig:elemental-fractionation} is that, as expected, noble gases fractionate according to their molecular weight. It can be seen that \ce{He}, the lightest noble gas, experienced the most depletion compared to its freestream value, while \ce{Xe}, the heaviest noble gas, experienced the most enrichment compared to its freestream value. Moreover, the species with a molecular weight lower than the molecular weight of the mixture in the tank get depleted, while the species with molecular weights larger get enriched.
Furthermore, the same conclusion applies to isotopes of a particular noble gas. They will also fractionate due to their molecular weight, with the heaviest isotopes being enriched, as can also be seen on Fig.~\ref{fig:elemental-fractionation} (isotopic fractionation). It should be noted that the fit shown in Fig.~\ref{fig:elemental-fractionation} in red and in Fig.~\ref{fig:isotopic-ratios-plot} in blue is for illustrative purposes only, and demonstrates that fractionation is mass-dependent. If a specific value of fractionation is needed for a given set of initial conditions, then a dedicated set of DSMC simulations would be used. The fit is also useful for performing efficient preliminary calculations/estimates related to the performance of the sampling system.

Since various isotopes of the same noble gas atom are modeled with the same transport properties except for their masses, only the mass of a noble gas isotope should be a factor in its fractionation compared to another isotope. We now switch from mass fractions to mole fractions to be consistent with how the planetary science community reports isotopic ratios. Fig.~\ref{fig:isotopic-ratios-plot} shows the mole fraction ratio of the various isotopes of \ce{Xe} when compared to $^{130}$\ce{Xe}, various isotopes of \ce{Kr} when compared to $^{84}$\ce{Kr}, and of various isotopes \ce{Ar} when compared to $^{36}$\ce{Ar}, as well as the associated error bars for each ratio, which represent one standard deviation from the average based on the DSMC results when averaged/sampled temporally. The blue lines represent the fit from eqn.~\ref{eq:ele-frac-fit}.
Furthermore, Table~\ref{tab:isotopic-ratios-tab} provides the isotopic ratios as well as one standard deviation over the average (also known as the relative standard deviation, or RSD, representative of the temporal fluctuations in \,\%) for various isotope pairs that are of interest to the Venus scientific community. Each data point and associated error bar in Fig.~\ref{fig:isotopic-ratios-plot} represent results from an individual DSMC simulation.

\begin{table*}
    \centering
    \caption{Isotopic fractionation ratios for various isotope pairs of interest.}\vspace{1mm}
    \resizebox{\textwidth}{!}{\begin{tabular}{ccccccccccccccc}
            \hline\hline
            Noble gas & Xe & Xe & Xe & Xe & Xe & Xe & Xe & Xe & He & Ne & Ar & Ar & Kr & Kr \\
            \hline
            isotope pair  & 128/130 & 129/130 & 132/130 & 136/130 & 128/132 & 129/132 & 130/132 & 136/132 & 3/4     & 20/22   & 38/36   & 40/36   & 82/84   & 86/84   \\
            \hline
            $\mu$          & \num{0.99892} & \num{0.99928} & \num{1.00088} & \num{1.00534} & \num{0.99799} & \num{0.99840} & \num{0.99912} & \num{1.00451} & \num{0.83800} & \num{0.93150} & \num{1.03409} & \num{1.06845} & \num{0.99221} & \num{1.00897} \\
            $\sigma$/$\mu$ & \num[round-mode=places, round-precision=2]{0.05}\,\%  & \num[round-mode=places, round-precision=2]{0.03}\,\%  & \num[round-mode=places, round-precision=2]{0.04}\,\%  & \num[round-mode=places, round-precision=2]{0.06}\,\%  & \num[round-mode=places, round-precision=2]{0.05}\,\%  & \num[round-mode=places, round-precision=2]{0.06}\,\%  & \num[round-mode=places, round-precision=2]{0.04}\,\%  & \num[round-mode=places, round-precision=2]{0.07}\,\%  & \num[round-mode=places, round-precision=2]{0.12}\,\%  & \num[round-mode=places, round-precision=2]{0.09}\,\%  & \num[round-mode=places, round-precision=2]{0.05}\,\%  & \num[round-mode=places, round-precision=2]{0.09}\,\%  & \num[round-mode=places, round-precision=2]{0.05}\,\%  & \num[round-mode=places, round-precision=2]{0.04}\,\%  \\
        \end{tabular}}
    \label{tab:isotopic-ratios-tab}
\end{table*}
\npnoround

It can be seen that the temporal fluctuations of the results from the DSMC simulations for each of those isotope pairs is always under 0.1\,\%, except for \ce{He}, which is due to the fast diffusion speed of \ce{He} because of its extremely light molar mass when compared with the average mixture molar mass. Additionally, as predicted previously, heavier isotopes are always enriched when compared to lighter isotopes.

\paragraph{Effect of statistics.}
\label{sec:statistics}

Inherently, DSMC is a stochastic method that represents a large amount of real molecules by a small number of simulated molecules, and therefore requires time averaging to reduce noise caused by statistical fluctuations. All collision procedures use an acceptance/rejection algorithm, and make use of a random number generator. To perform statistically-independent simulations, SPARTA allows the user to choose a single seed to initialize the random number generator. Therefore, by performing multiple simulations with different initial seeds, one can quantify the effects of statistics. We performed five identical simulations that only differed by their initial seed, for all fifteen noble gases, and at reduced density (-25\,\%) conditions. We chose the reduced density condition because of its significantly lower computational cost to run the 75 required simulations. It was found that the RSD was under 0.2\,\% for all fifteen noble gases, except for $^{4}$\ce{He}, for which it was under 0.4\,\%. Table~\ref{tab:minus25-density} shows the normalized mass fraction for all five random seeds as well as the average value and relative standard deviation for the fifteen noble gas species. Given the low amount of statistical fluctuations between the runs, we are confident that our DSMC simulations are properly resolved with respect to the number of computational particles, which is a critical aspect of the DSMC method.

\begin{table*}
    \centering
    \caption{Normalized mass fractions for the fifteen noble gas species, at -25\,\% density freestream conditions.}\vspace{1mm}
    \resizebox{\textwidth}{!}{\begin{tabular}{cccccccccccccccc}
            \hline\hline
            Noble gas & $^{3}$He & $^{4}$He & $^{20}$Ne & $^{22}$Ne & $^{36}$Ar & $^{38}$Ar & $^{40}$Ar & $^{82}$Kr & $^{84}$Kr & $^{86}$Kr & $^{128}$Xe & $^{129}$Xe & $^{130}$Xe & $^{132}$Xe & $^{136}$Xe \\
            \hline
            Seed1               & \num{0.17503}  & \num{0.21291}  & \num{0.79482}   & \num{0.86321}   & \num{1.32653}   & \num{1.37734}   & \num{1.43057}   & \num{2.08305}   & \num{2.08986}   & \num{2.10976}   & \num{2.18345}    & \num{2.18415}    & \num{2.18500}    & \num{2.19747}    & \num{2.19838}    \\
            Seed2               & \num{0.17537}  & \num{0.21301}  & \num{0.79570}   & \num{0.86476}   & \num{1.32527}   & \num{1.37695}   & \num{1.42922}   & \num{2.07710}   & \num{2.09982}   & \num{2.10919}   & \num{2.18488}    & \num{2.18343}    & \num{2.18581}    & \num{2.18955}    & \num{2.19472}    \\
            Seed3               & \num{0.17523}  & \num{0.21166}  & \num{0.79471}   & \num{0.86065}   & \num{1.32981}   & \num{1.38023}   & \num{1.43013}   & \num{2.07838}   & \num{2.09646}   & \num{2.11048}   & \num{2.18066}    & \num{2.18493}    & \num{2.18875}    & \num{2.18887}    & \num{2.19709}    \\
            Seed4               & \num{0.17489}  & \num{0.21194}  & \num{0.79389}   & \num{0.86091}   & \num{1.32941}   & \num{1.37712}   & \num{1.42756}   & \num{2.08190}   & \num{2.09628}   & \num{2.10502}   & \num{2.18114}    & \num{2.18264}    & \num{2.18617}    & \num{2.18752}    & \num{2.19294}    \\
            Seed5               & \num{0.17459}  & \num{0.21110}  & \num{0.79412}   & \num{0.86318}   & \num{1.32505}   & \num{1.37786}   & \num{1.42838}   & \num{2.08059}   & \num{2.09932}   & \num{2.10778}   & \num{2.18418}    & \num{2.18157}    & \num{2.18440}    & \num{2.18770}    & \num{2.19742}    \\
            \hline
            $\mu$               & \num{0.17502}  & \num{0.21212}  & \num{0.79465}   & \num{0.86254}   & \num{1.32721}   & \num{1.37790}   & \num{1.42917}   & \num{2.08020}   & \num{2.09635}   & \num{2.10844}   & \num{2.18286}    & \num{2.18335}    & \num{2.18603}    & \num{2.19022}    & \num{2.19611}    \\
            $\sigma$/$\mu$      & \num[round-mode=places, round-precision=2]{0.173}\,\%  & \num[round-mode=places, round-precision=2]{0.386}\,\%  & \num[round-mode=places, round-precision=2]{0.089}\,\%   & \num[round-mode=places, round-precision=2]{0.201}\,\%   & \num[round-mode=places, round-precision=2]{0.170}\,\%   & \num[round-mode=places, round-precision=2]{0.098}\,\%   & \num[round-mode=places, round-precision=2]{0.086}\,\%   & \num[round-mode=places, round-precision=2]{0.118}\,\%   & \num[round-mode=places, round-precision=2]{0.189}\,\%   & \num[round-mode=places, round-precision=2]{0.102}\,\%   & \num[round-mode=places, round-precision=2]{0.086}\,\%    & \num[round-mode=places, round-precision=2]{0.060}\,\%    & \num[round-mode=places, round-precision=2]{0.076}\,\%    & \num[round-mode=places, round-precision=2]{0.189}\,\%    & \num[round-mode=places, round-precision=2]{0.101}\,\%    \\
        \end{tabular}}
    \label{tab:minus25-density}
\end{table*}

\paragraph{Effect of surface boundary conditions.}
\label{sec:surfaceBC}

As mentioned in Sec.~\ref{subsec:simulationsparameters}, we performed simulations with both a cold wall boundary condition where the surface temperature is fixed at 295~K, as well as a hot wall boundary condition which uses the radiative equilibrium law previously described. The latter leads to a peak surface temperature $\approx$~\num{2100}~K near the entrance of the sampling inlet. However, we notice that the temperature inside the inlet drops rapidly, to under \num{1000}~K in a matter of a few millimeters. The freestream number density, velocity and maximum level of grid refinement are all kept constant at the baseline value. Table~\ref{tab:surfaceBC-comparison} shows the comparison of the normalized mass fraction for all fifteen noble gas species for both surface boundary conditions. It can be seen that the differences between both surface boundary conditions are under 1\,\% for all cases, and on average around a half a percent, which can be considered on the order of the statistical fluctuations. Moreover, the heavier noble gases are slightly more affected than lighter ones.

\begin{table*}
    \centering
    \caption{Normalized mass fraction for the fifteen noble gas species, at baseline freestream conditions, for cold and hot (using the re-radiative surface boundary condition) walls. Both sets of simulations use a maximum grid refinement level of 10.}\vspace{1mm}
    \resizebox{\textwidth}{!}{\begin{tabular}{cccccccccccccccc}
            \hline\hline
            Noble gas & $^{3}$He & $^{4}$He & $^{20}$Ne & $^{22}$Ne & $^{36}$Ar & $^{38}$Ar & $^{40}$Ar & $^{82}$Kr & $^{84}$Kr & $^{86}$Kr & $^{128}$Xe & $^{129}$Xe & $^{130}$Xe & $^{132}$Xe & $^{136}$Xe \\
            \hline
            Cold wall (ref.) & \num{0.21454}  & \num{0.25766}  & \num{0.85844}   & \num{0.92126}   & \num{1.33030}   & \num{1.37341}   & \num{1.41831}   & \num{2.00172}   & \num{2.01281}   & \num{2.03040}   & \num{2.10951}    & \num{2.11102}    & \num{2.11045}    & \num{2.11281}    & \num{2.11631}    \\
            Hot wall     & \num{0.21393}  & \num{0.25764}  & \num{0.86088}   & \num{0.92258}   & \num{1.33536}   & \num{1.37905}   & \num{1.42575}   & \num{2.01102}   & \num{2.02603}   & \num{2.04386}   & \num{2.12853}    & \num{2.12651}    & \num{2.12647}    & \num{2.13010}    & \num{2.13343}    \\
            \hline
            Relative difference & \num[round-mode=places, round-precision=2]{0.286}\,\%  & \num[round-mode=places, round-precision=2]{0.008}\,\%  & \num[round-mode=places, round-precision=2]{-0.285}\,\%  & \num[round-mode=places, round-precision=2]{-0.143}\,\%  & \num[round-mode=places, round-precision=2]{-0.380}\,\%  & \num[round-mode=places, round-precision=2]{-0.411}\,\%  & \num[round-mode=places, round-precision=2]{-0.524}\,\%  & \num[round-mode=places, round-precision=2]{-0.464}\,\%  & \num[round-mode=places, round-precision=2]{-0.657}\,\%  & \num[round-mode=places, round-precision=2]{-0.663}\,\%  & \num[round-mode=places, round-precision=2]{-0.902}\,\%   & \num[round-mode=places, round-precision=2]{-0.734}\,\%   & \num[round-mode=places, round-precision=2]{-0.759}\,\%   & \num[round-mode=places, round-precision=2]{-0.819}\,\%   & \num[round-mode=places, round-precision=2]{-0.809}\,\% \\
        \end{tabular}}
    \label{tab:surfaceBC-comparison}
\end{table*}

\paragraph{Effect of freestream number density.}
\label{sec:density}

We then performed sensitivity studies on the freestream number density and its impact on fractionation. In that order, aside from the baseline density, we also considered conditions where the density was 25\,\% higher and 25\,\% lower. There is still significant uncertainty associated with predicting Venus atmospheric conditions at the sampling altitude, along with expected diurnal variations~\citep{ref:mahieux2012}. Varying density by 25\,\% in each direction represents a sizable variation, yet also balances computational cost associated with increasing the freestream density. It is important to note that we did not vary the maximum grid refinement level of 10 in any of those simulations. Since the mean free path for a VSS gas is inversely proportional to the number density, the grid cell size should ordinarily be reduced when the number density increases. However, since our simulations were capped, for this sensitivity study on the freestream number density, we kept the maximum grid refinement level fixed at 10, which is the maximum number that is computationally acceptable.
Table~\ref{tab:density-comparison} shows the comparison of the normalized mass fraction for all fifteen noble gas species for the various density conditions. Interesting trends arise based on the mass of the noble gas of interest. Firstly, the lighter gas \ce{He} is the most affected by density changes, in particular its lighter isotope $^{3}$He. Additionally, the lighter noble gases are enriched when the density decreases, and depleted when the density increases, but the trend is reversed for all noble gases heavier than $^{22}$Ne.

\begin{table*}
    \centering
    \caption{Normalized mass fraction for the fifteen noble gas species, at baseline freestream conditions, except for +25\,\% and -25\,\% variations in the number density. All sets of simulations use a maximum grid refinement level of 10.}\vspace{1mm}
    \resizebox{\textwidth}{!}{\begin{tabular}{cccccccccccccccc}
            \hline\hline
            Noble gas  & $^{3}$He  & $^{4}$He  & $^{20}$Ne & $^{22}$Ne & $^{36}$Ar & $^{38}$Ar & $^{40}$Ar & $^{82}$Kr & $^{84}$Kr & $^{86}$Kr & $^{128}$Xe & $^{129}$Xe & $^{130}$Xe & $^{132}$Xe & $^{136}$Xe \\
            \hline
            Nominal num. density (ref.) & \num{0.21454}   & \num{0.25766}   & \num{0.85844}   & \num{0.92126}   & \num{1.33030}   & \num{1.37341}   & \num{1.41831}   & \num{2.00172}   & \num{2.01281}   & \num{2.03040}   & \num{2.10951}    & \num{2.11102}    & \num{2.11045}    & \num{2.11281}    & \num{2.11631}    \\
            \hline
            -25\,\% num. density  & \num{0.17503}   & \num{0.21290}   & \num{0.79482}   & \num{0.86321}   & \num{1.32653}   & \num{1.37734}   & \num{1.43057}   & \num{2.08305}   & \num{2.08986}   & \num{2.10976}   & \num{2.18345}    & \num{2.18415}    & \num{2.18500}    & \num{2.19747}    & \num{2.19838}    \\
            Relative difference  & \num[round-mode=places, round-precision=2]{18.418}\,\%  & \num[round-mode=places, round-precision=2]{17.370}\,\%  & \num[round-mode=places, round-precision=2]{7.410}\,\%   & \num[round-mode=places, round-precision=2]{6.301}\,\%   & \num[round-mode=places, round-precision=2]{0.283}\,\%   & \num[round-mode=places, round-precision=2]{-0.287}\,\%  & \num[round-mode=places, round-precision=2]{-0.864}\,\%  & \num[round-mode=places, round-precision=2]{-4.063}\,\%  & \num[round-mode=places, round-precision=2]{-3.828}\,\%  & \num[round-mode=places, round-precision=2]{-3.909}\,\%  & \num[round-mode=places, round-precision=2]{-3.505}\,\%   & \num[round-mode=places, round-precision=2]{-3.464}\,\%   & \num[round-mode=places, round-precision=2]{-3.532}\,\%   & \num[round-mode=places, round-precision=2]{-4.007}\,\%   & \num[round-mode=places, round-precision=2]{-3.878}\,\%   \\
            \hline
            +25\,\% num. density  & \num{0.25554}   & \num{0.30401}   & \num{0.90379}   & \num{0.96293}   & \num{1.33271}   & \num{1.37720}   & \num{1.41127}   & \num{1.97691}   & \num{1.96901}   & \num{2.700624}  & \num{2.08752}    & \num{2.07998}    & \num{2.08533}    & \num{2.08941}    & \num{2.08978}    \\
            Relative difference  & \num[round-mode=places, round-precision=2]{-19.112}\,\% & \num[round-mode=places, round-precision=2]{-17.991}\,\% & \num[round-mode=places, round-precision=2]{-5.284}\,\%  & \num[round-mode=places, round-precision=2]{-4.524}\,\%  & \num[round-mode=places, round-precision=2]{-0.181}\,\%  & \num[round-mode=places, round-precision=2]{-0.276}\,\%  & \num[round-mode=places, round-precision=2]{0.496}\,\%   & \num[round-mode=places, round-precision=2]{1.240}\,\%   & \num[round-mode=places, round-precision=2]{2.176}\,\%   & \num[round-mode=places, round-precision=2]{1.190}\,\%   & \num[round-mode=places, round-precision=2]{1.042}\,\%    & \num[round-mode=places, round-precision=2]{1.470}\,\%    & \num[round-mode=places, round-precision=2]{1.190}\,\%    & \num[round-mode=places, round-precision=2]{1.107}\,\%    & \num[round-mode=places, round-precision=2]{1.253}\,\%    \\
        \end{tabular}}
    \label{tab:density-comparison}
\end{table*}

\paragraph{Effect of freestream velocity.}
\label{sec:velocity}

Subsequently, we performed a study where the freestream velocity was changed to 13.0~km/s, which corresponds to a Mach number of 61.3. That value was chosen because in an initial iteration of the mission concept, VATMOS-SR was baselined as a secondary payload to the Dragonfly trajectory, that included a Venus flyby, and the probe would have been skimming the atmosphere at $\approx13.0$~km/s.
For this particular study, we used the radiative equilibrium wall boundary condition, since, we found that at 10.5~km/s, the differences between the cold wall and radiative equilibrium wall boundary conditions were minimal, and also because we expect the surface temperature to be much higher at a freestream velocity of 13.0~km/s.
As can be seen in Table~\ref{tab:velocity-comparison}, the increase in the freestream velocity has a similar effect to a decrease in the density, that is, it depletes the lighter noble gases and enriches the heavier ones. However, this time, the \say{tipping point} is located between $^{20}$Ne and $^{22}$Ne, around a molecular weight of 21~g/mol. That can be explained by the fact that the velocity increase leads to a larger of amount of dissociation, and therefore decreases the mixture average molecular weight (by 0.85~g/mol in the tank).

\begin{table*}
    \centering
    \caption{Normalized mass fraction for the fifteen noble gas species, at baseline freestream conditions, except for a variation in the velocity. Both sets of hot wall simulations use the radiative equilibrium wall boundary condition and a maximum grid refinement level of 10. The relative difference is between the two hot wall cases.}\vspace{1mm}
    \resizebox{\textwidth}{!}{\begin{tabular}{cccccccccccccccc}
            \hline\hline
            Noble gas & $^{3}$He  & $^{4}$He  & $^{20}$Ne & $^{22}$Ne & $^{36}$Ar & $^{38}$Ar & $^{40}$Ar & $^{82}$Kr & $^{84}$Kr & $^{86}$Kr & $^{128}$Xe & $^{129}$Xe & $^{130}$Xe & $^{132}$Xe & $^{136}$Xe \\
            \hline
            Cold wall 10.5~km/s & \num{0.21454}   & \num{0.25766}   & \num{0.85844}   & \num{0.92126}   & \num{1.33030}   & \num{1.37341}   & \num{1.41831}   & \num{2.00172}   & \num{2.01281}   & \num{2.03040}   & \num{2.10951}    & \num{2.11102}    & \num{2.11045}    & \num{2.11281}    & \num{2.11631}    \\ \hline
            Hot wall 10.5~km/s (ref.) & \num{0.21393}   & \num{0.25764}   & \num{0.86088}   & \num{0.92258}   & \num{1.33536}   & \num{1.37905}   & \num{1.42575}   & \num{2.01102}   & \num{2.02603}   & \num{2.04386}   & \num{2.12853}    & \num{2.12651}    & \num{2.12647}    & \num{2.13010}    & \num{2.13343}    \\
            Hot wall 13.0 km/s & \num{0.26366}   & \num{0.31554}   & \num{0.98259}   & \num{1.04783}   & \num{1.43689}   & \num{1.48212}   & \num{1.52026}   & \num{1.98599}   & \num{1.99457}   & \num{2.01631}   & \num{1.99581}    & \num{1.99897}    & \num{2.00021}    & \num{2.00728}    & \num{2.00281}    \\
            \hline
            Relative difference & \num[round-mode=places, round-precision=2]{-23.247}\,\% & \num[round-mode=places, round-precision=2]{-22.473}\,\% & \num[round-mode=places, round-precision=2]{-14.137}\,\% & \num[round-mode=places, round-precision=2]{-13.577}\,\% & \num[round-mode=places, round-precision=2]{-7.603}\,\%  & \num[round-mode=places, round-precision=2]{-7.474}\,\%  & \num[round-mode=places, round-precision=2]{-6.629}\,\%  & \num[round-mode=places, round-precision=2]{1.245}\,\%   & \num[round-mode=places, round-precision=2]{1.553}\,\%   & \num[round-mode=places, round-precision=2]{1.348}\,\%   & \num[round-mode=places, round-precision=2]{6.235}\,\%    & \num[round-mode=places, round-precision=2]{5.998}\,\%    & \num[round-mode=places, round-precision=2]{5.938}\,\%    & \num[round-mode=places, round-precision=2]{5.766}\,\%    & \num[round-mode=places, round-precision=2]{6.123}\,\% \\
        \end{tabular}}
    \label{tab:velocity-comparison}
\end{table*}

\section{Discussion}
\label{sec:discussion}

In the previous section, we performed 3-dimensional simulations of the VATMOS-SR spacecraft, and showed that the stagnation pressure reached $\approx 380$~Pa in the sampling inlet. The lightest and heaviest noble gases, $^{3}$\ce{He} and $^{136}$\ce{Xe}, showed opposite behaviors in the bow shock and compression layer regions. The strong compression layer encountered behind a Mach 50 shock depletes the lightest gases and enriches the heaviest ones. We also noticed that despite the pressure reaching a steady-state and constant value throughout the length of the first sampling inlet prior to the valve opening, the normalized mass fraction of the noble gases varied spatially along the stagnation streamline. By comparing with results for the normal shock from Appendix~\ref{sec:fractionation}, we can conclude that the pressure diffusion effects have a much stronger impact than the thermal diffusion effects, and are the driving factor for fractionation in this blunt-body flow. Interestingly, as the strong pressure gradient in front of the vehicle drives the fractionation of the sample (and not the valve or internal plumbing), this suggests that the elemental composition of the gas mixture that enters the sampling system is comparable to that of the elemental composition in the sample tanks. Therefore, the complex Mindrum valve geometry may not be required in the computational domain to accurately predict fractionation, and a simplified geometry could likely be used instead, in the future, with comparable accuracy.

Subsequently, motivated by the science objectives of VATMOS-SR, we investigated the effects of various parameters on our main quantity of interest, the mass fraction of noble gases collected in the sampling tank normalized by the mass fraction of noble gases in the freestream. We studied both elemental fractionation, which quantifies how different noble gases fractionate, and isotopic fractionation, which quantifies how two isotopes of the same noble gas fractionate. The latter is an important metric, since, in our simulations, the only difference between two isotopes of the same noble gases is their mass, and therefore offers a direct view into the impact of mass on differential diffusion.

The transfer function that we computed relates each noble gas' sampled mass fraction with their relative freestream mass fraction, and was fitted from data obtained independently for each noble gas species using a one-phase exponential association equation based on a standard Weibull function. Since $^{136}$\ce{Xe} gets enriched and $^{3}$\ce{He} depleted, the former has a number density roughly an order of magnitude higher than the latter, which also means an order of magnitude more simulated molecules, $N$. Since statistical fluctuations correlate with $1/\sqrt{N}$, we expect the statistical fluctuation to be significantly lower on \ce{Xe} than on \ce{He}. That is one possible explanation as to why the Weibull function fits the \ce{Xe} data better than the \ce{He} data.

Our investigations showed that the statistics inherently associated with the DSMC method, that are a by-product of the use of random numbers, lead to fluctuations that are on average under 0.2\,\%. We can therefore consider that our simulations are well resolved in terms of the number of particles. The last numerical parameter we investigated was the surface boundary condition, and we observed that the difference between using cold wall and re-radiative boundary conditions led to relative differences that were on average under 0.7\,\%.

We then investigated input parameters that are related to the Venusian atmosphere and mission concept, the freestream number density and velocity, respectively. Those two parameters were found to more substantially affect the normalized mass fraction of the noble gases. Starting with the number density, we added uncertainties of $\pm25$\,\% around our nominal value, to account for potential unexpected variations in the Venusian atmospheric density as well as the sampling altitude. We then focused on the effects of a change in the freestream velocity from 10.5 to 13.0~km/s, which could be a realistic change depending on the launch vehicle and overall mission concept. Our studies reported a few important conclusions, which are presented henceforth along with their interpretations:
\begin{enumerate}
    \item \textbf{Lighter noble gases are more affected by changes in the freestream density and velocity than heavier ones:} The velocity of lighter atoms is significantly affected by collisions with all other molecules, whereas, for heavier atoms, their velocity is only significantly affected by collision with molecules of comparable mass.
    \item \textbf{Lighter noble gases are enriched when the density or velocity decreases, and depleted when the density or velocity increases. The opposite is true for heavier noble gases:} We previously concluded that regions of higher pressure enriched the heavier noble gases, and depleted the lighter ones. A stronger shock or higher velocity would have the effect of increasing the stagnation pressure, therefore enriching the heavier noble gases.
\end{enumerate}

The latter results are important because they showcase that possible uncertainties in the Venusian atmospheric composition during the atmospheric pass may have a large impact on the measurement of the abundance of noble gases. Is it therefore crucial that any Venus exploration missions that could launch before or concurrently with VATMOS-SR, such as VERITAS~\citep{Smrekar2022}, EnVision~\citep{ghail2017envision}, and DAVINCI~\citep{garvin2022}, could provide data and analysis that would yield a higher confidence in our ability to predict the Venusian atmospheric composition and thermodynamic properties. Furthermore, precise knowledge of the freestream density and temperature and sampling speed for VATMOS-SR will be required \emph{a posteriori} for accurate post-flight simulations analysis. Relating this analysis and discussion to the scientific objectives of VATMOS-SR, an isotopic fractionation uncertainty lower than 1\%/atomic mass unit (amu) of the sampled gas would still allow determining the extent of atmospheric loss of atmospheric xenon on Venus compared to the case of Earth and Mars (40\%/amu). It would also enable determining if comets delivered volatile elements to Venus. Having such limited isotopic fractionation also means that excesses of noble gas isotopes produced by the decay of radioactive elements (e.g., $^{129}$\ce{I}, $^{238}$\ce{U}, $^{40}$\ce{K}) would still be detectable~\cite{avice_noble_2022}. As shown in Fig.~\ref{fig:isotopic-ratios-plot}, the overall isotopic fraction is significantly less than 1\%/amu, and we have also demonstrated the ability to predict this fractionation with DSMC simulations. Future work will continue to quantitatively relate the uncertainty associated with predicting fractionation to the scientific objectives of VATMOS-SR.

It is worth noting that one critical missing parameter is the knowledge of how to model different isotopes of the same noble gas. The approach chosen in this work was to simply change the mass of each isotope, but this assumption was driven as much by the lack of any data associated with isotopic specific transport properties, as it was by the fact that isotopes of the same species have the same electron cloud and therefore the same interatomic potential and VSS parameters. Experimental studies measuring the diffusion coefficient of several isotopes of a noble gas would be crucially important to provide validation for this work, as they could then potentially open the door to fitting different VSS parameters to isotopes of the same species.

While many parameters were varied in this work, it is not meant to represent an exhaustive sensitivity study due to the complexity of the topic being studied. We briefly highlight several areas that could be the topic of future work. No gas--surface interactions were included in this work due to the uncertainty associated with gas--surface interaction models, along with the uncertainty with materials that will be used in VATMOS-SR spacecraft. Similarly, only one thermal boundary condition was used in this study. When additional detail is available, investigating the sensitivity of fractionation to gas--surface interaction modeling choices along with thermal surface boundary conditions is a logical course of action to pursue. The effect of grid refinement should also be investigated in more detail for the higher-density cases that were included, where it was not possible to spatially resolve the mesh down to the mean free path of the flow in certain areas of the domain. Furthermore, simulations in this work assumed a 0$^\circ$ angle of attack (AoA), and future parameter space investigations could include simulations with non-zero AoAs based on expected vehicle flight dynamics. However, these simulations will be computationally more expensive due to decreased symmetry with the vehicle geometry. If the VATMOS-SR mission concept is selected for further study, we expect additional work to be performed to quantify the sensitivity of the fractionation predictions from numerical simulations. 

Due to the rarefied high-speed flow regime of interest, end-to-end experimental validation of fractionation predicted by DSMC simulations at flight-relevant conditions through the entire VATMOS-SR sampling system is likely impossible. However, future work should investigate the feasibility of experimentally validating different aspects of the computations/sampling system in a piecewise manner. For example, once TPS material selection is finalized, experimental work could be performed to investigate if ablation products in any significant quantity are collected in the sampling tanks, in addition to the Venus atmospheric sample. 

Overall, we have demonstrated the use of DSMC simulations with a complex flight-relevant geometry to predict elemental and isotopic fractionation for the VATMOS-SR mission concept.
\textbf{Based on these simulations, we are confident that the freestream concentration is uniquely correlated to concentration in the tanks, with a good degree of certainty.}
These predictions will be important for VATMOS-SR mission concept to relate the mixture composition of the acquired Venus gas samples to the true atmospheric composition of the Venus atmosphere.


\section*{Acknowledgements}
The authors would like to thank Stan Moore (Sandia National Laboratories) and Steve Plimpton (retired, Sandia National Laboratories) for their help implementing new SPARTA features that were necessary to perform this work.
Resources supporting this work were provided by the NASA High-End Computing (HEC) Program through the NASA Advanced Supercomputing (NAS) Division at Ames Research Center.
Parts of this work have been performed at the Jet Propulsion Laboratory (JPL), California Institute of Technology, under contract to NASA.
Arnaud Borner was partly funded by JPL subcontract No. 1656387 to Analytical Mechanics Associates, Inc.
Sandia National Laboratories is a multi-mission laboratory managed and operated by National Technology and Engineering Solutions of Sandia, LLC, a wholly owned subsidiary of Honeywell International, Inc., for the U.S. Department of Energy’s National Nuclear Security Administration under contract DE-NA0003525.
Guillaume Avice thanks Prof. Ken Farley for his support during his postdoctoral scholarship at the California Institute of Technology. Guillaume Avice and Christophe Sotin thank the french Centre National d'Etudes Spatiales (CNES) for its support to the VATMOS-SR mission concept.

\section*{Data Availability}
The results in this study were generated using SPARTA, available on the \href{https://github.com/sparta/sparta}{SPARTA Github page}~\citep{plimpton2019direct}. The input decks are available from the corresponding author upon reasonable request and pending export considerations.

\appendix
\section{VSS Collision Model Parameters} 
\label{sec:collisionmodelparameters}

The transport properties in DSMC are not directly provided as input nor computed from collision integrals (CI). They are a direct result of the collisional interactions between coarse-grained DSMC particles. The outcome of these particle collisions is determined by the collision model employed and the user-provided parameters. Hence, the accuracy of transport properties within DSMC is directly dependent upon the input collision model parameters.

When applying the VSS model to collisions between dissimilar species, an important aspect is how the model parameters are specified. One common method, known as the \emph{collision-averaged} approach, treats the collision parameters as intrinsic to each species, irrespective of the species it collides with. In this framework, parameters for interactions between different species are obtained by averaging the individual species’ values. Alternatively, the \emph{collision-specific} approach assigns unique DSMC parameters to each pair of colliding species, allowing for more tailored modeling of inter-species interactions.

The \emph{collision-averaged} values for the reacting species that make up the Venusian atmosphere are obtained from Swaminathan-Gopalan~\citep{swaminathan2015consistent} (no data was available for \ce{C2}, therefore, the same parameters as \ce{CO} are used, due to the similarities between the two molecules). The values for the noble gases are obtained from various references~\citep{swaminathan2016recommended,bird1994molecular,weaver2015revised}.

On the other hand, for the \emph{collision-specific} approach, the interaction between each of the 15 species in the mixture has to be considered.
Therefore, the methodology described by Swaminathan-Gopalan and Stephani~\citep{swaminathan2016recommended} involving the fitting of VSS parameters using the CI is employed here.
The CI data for some of these interactions were readily available~\citep{kestin1984equilibrium,bzowski1990equilibrium}. However, the CI data for other interactions were not present in the literature to the best of the authors' knowledge.
For these, the CIs are computed using the Lennard-Jones (LJ) potential from the code developed by Subramaniam~\citep{subramaniam2020state}. Although the LJ potential is not highly accurate, the occurrence of these collisions is quite rare, and hence this level of accuracy is sufficient for the current study. The LJ parameters are taken from the work of Rutkai~\citep{rutkai2017well}.
Once we have all the collision integrals, we use the methodology described in Ref.~\citep{swaminathan2016recommended} to obtain the collision model parameters by fitting to the CI data, which are then used in the DSMC simulations. 

We compared the \emph{collision-averaged} and \emph{collision-specific} approaches and found that, for our problem of interest, the discrepancies between the two approaches were on the order of the statistical fluctuations shown in Table~\ref{tab:minus25-density}, and can therefore be considered negligible. Furthermore, a recent publication~\citep{hong2023optimized} suggested that \emph{collision-averaged} were more accurate than \emph{collision-specific} parameters near ambient conditions. The \emph{collision-averaged} VSS parameters were therefore used for all results shown in this study, and are given in Table~\ref{tab:VSS_params}.

\begin{table*}
	\centering
	\caption{VSS collision model parameters. $d_{ref}$ is the reference diameter, $\omega$ is the temperature exponent, $\alpha$ is the scattering exponent, and $T_{ref}$ is the reference temperature.}
		\begin{tabular}{cccccc}
			\hline
            \hline
			Species & $d_{ref}$ [\AA] & $\omega$ & $\alpha$ & $T_{ref}$ (K) & Ref.\\
			\hline
\ce{CO2} & 4.147 & 0.632 & 1.259 & 273 & \citep{swaminathan2015consistent}, Table 5.10\\
\ce{CO}  & 4.684 & 0.787 & 1.494 & 273 & \citep{swaminathan2015consistent}, Table 5.10\\
\ce{O2}  & 3.896 & 0.700 & 1.463 & 273 & \citep{swaminathan2015consistent}, Table 5.3\\
\ce{N2}  & 4.040 & 0.686 & 1.424 & 273 & \citep{swaminathan2015consistent}, Table 5.3\\
\ce{NO}  & 4.218 & 0.737 & 1.542 & 273 & \citep{swaminathan2015consistent}, Table 5.3\\
\ce{CN}  & 4.609 & 0.784 & 1.635 & 273 & \citep{swaminathan2015consistent}, Table 5.10\\
\ce{C2}  & 4.684 & 0.787 & 1.494 & 273 & \citep{swaminathan2015consistent}, Table 5.10\\
\ce{O}   & 3.692 & 0.803 & 1.582 & 273 & \citep{swaminathan2015consistent}, Table 5.3\\
\ce{N}   & 3.697 & 0.790 & 1.486 & 273 & \citep{swaminathan2015consistent}, Table 5.10\\
\ce{C}   & 4.983 & 0.883 & 1.523 & 273 & \citep{swaminathan2015consistent}, Table 5.10\\
			\hline
\ce{He}  & 2.298 & 0.712 & 1.429 & 350    & \citep{weaver2015revised}, Table 3\\
\ce{Ne}  & 2.484 & 0.675 & 1.331 & 823.15 & \citep{weaver2015revised}, Table 3\\
\ce{Ar}  & 4.011 & 0.722 & 1.470 & 273    & \citep{swaminathan2016recommended}, Table 3\\
\ce{Kr}  & 4.096 & 0.734 & 1.312 & 473.15 & \citep{weaver2015revised}, Table 3\\
\ce{Xe}  & 5.650 & 0.850 & 1.440 & 273    & \citep{bird1994molecular}, Tables A1 \& A3\\
			\hline
		\end{tabular}
    \label{tab:VSS_params}
\end{table*}  

\section{TCE Chemical Reactions Rates} 
\label{sec:chemical_rates}

The reaction rates used in the TCE model in SPARTA are presented in Table~\ref{TCErates_dissrecomb} for dissociation and their corresponding recombination reactions, and Table~\ref{TCErates_exchange} for all exchange reactions, respectively. Since SPARTA does not currently allow for the calculation of backward reaction rates from specified forward reaction rates and equilibrium rates, both the forward and backward rates have to be specified. The backward rates were developed for this rate based on fitting the rates calculated by the ratio of the forward rate and the equilibrium constant computed from the Gibbs free energy to a modified Arrhenius rate. Furthermore, the TCE model in DSMC places additional constraints on the choice of the temperature exponent when fitting the modified Arrhenius rate, as explained by Higdon~\citep[Ch. 7]{higdon2018monte}.

\sisetup{round-mode=places, round-precision=3}
\begin{table*}
	\centering
	\caption{Chemical kinetics rates for dissociation and recombination reactions used in the TCE model. $E_{a}$ represents the Arrhenius activation energy, $A$ the Arrhenius pre-exponential factor, $n$ the modified Arrhenius temperature exponent and $\Delta H$ the reaction enthalpy. The $type$ represents Dissociation (D) and Recombination (R) reactions (see \citealp{ref:johnston2014,fujita2006impact,park1993review}).}
		\resizebox{\textwidth}{!}{\begin{tabular}{lllrrrrll}
            \hline
			\hline
			i & Type & Reaction & E$_a$ (J) & A (\unit{m^{3}.molecule^{-1}.s^{-1}}) & n & $\Delta$H (J) & Third Body, M & Ref. \\
            \hline
            1f & D & \ce{CO2 + M -> 2CO + O + M} & \num{8.736e-19} & \num{1.146e-9} & \num[round-mode=places, round-precision=1]{-1.5} & \num{-8.736e-19} & Noble gases & \citep{ref:johnston2014}\\
            & & & & \num{2.325e-8} & & & \ce{N,C,O} & \citep{ref:johnston2014}\\
            & & & & \num{1.146e-8} & & & Others & \citep{ref:johnston2014}\\
            1b & R & \ce{2CO + O + M -> CO2 + M} & \num[round-mode=places, round-precision=0]{0.0} & \num{3.614e-46} & \num{-0.238} & \num{8.736e-19} & Noble gases & This work\\
            & & & & \num{7.331e-45} & & & \ce{N,C,O} & This work\\
            & & & & \num{3.614e-45} & & & Others & This work\\
            2f & D & \ce{CO + M -> C + O + M} & \num{1.781e-18} & \num{1.986e-10} & \num[round-mode=places, round-precision=1]{-1.0} & \num{-1.781e-18} & Noble gases & \citep{ref:johnston2014}\\
            & & & & \num{2.936e-9} & & & \ce{N,C,O} & \citep{ref:johnston2014}\\
            & & & & \num{1.986e-9} & & & Others & \citep{ref:johnston2014}\\
            2b & R & \ce{C + O + M -> CO + M} & \num[round-mode=places, round-precision=0]{0.0} & \num{1.045e-41} & \num{-1.023} & \num{1.781e-18} & Noble gases & This work\\
            & & & & \num{1.542e-40} & & & \ce{N,C,O} & This work\\
            & & & & \num{1.045e-40} & & & Others & This work\\
            3f & D & \ce{C2 + M -> C + C + M} & \num{9.872e-19} & \num{7.473e-12} & \num[round-mode=places, round-precision=1]{-1.0} & \num{-9.872e-19} & All & \citep{ref:johnston2014}\\
            3b & R & \ce{C + C + M -> C2 + M} & \num[round-mode=places, round-precision=0]{0.0} & \num{1.545e-42} & \num{-0.421} & \num{9.872e-19} & All & This work\\
            4f & D & \ce{CN + M -> C + N + M} & \num{9.734e-19} & \num{9.963e-15} & \num[round-mode=places, round-precision=1]{-0.4} & \num{-9.734e-19} & All & \citep{fujita2006impact}\\
            4b & R & \ce{C + N + M -> CN + M} & \num[round-mode=places, round-precision=0]{0.0} & \num{1.690e-40} & \num[round-mode=places, round-precision=2]{-1.35} & \num{9.734e-19} & All & This work\\
            5f & D & \ce{N2 + M -> N + N + M} & \num{1.563e-18} & \num{4.982e-8} & \num[round-mode=places, round-precision=1]{-1.6} & \num{-1.563e-18} & \ce{N,C,O} & \citep{park1993review}\\
            & & & & \num{1.162e-8} & & & Others & \citep{park1993review}\\
            5b & R & \ce{N + N + M -> N2 + M} & \num[round-mode=places, round-precision=0]{0.0} & \num{4.937e-40} & \num[round-mode=places, round-precision=2]{-1.35} & \num{1.563e-18} & \ce{N,C,O} & This work\\
            & & & & \num{1.152e-40} & & & Others & This work\\
            6f & D & \ce{NO + M -> N + O + M} & \num{1.043e-18} & \num{7.448e-14} & \num[round-mode=places, round-precision=0]{0.0} & \num{-1.043e-18} & \ce{N,C,O,NO,CO2} & \citep{park1993review}\\
            & & & & \num{3.321e-15} & & & Others & \citep{park1993review}\\
            6b & R & \ce{N + O + M -> NO + M} & \num[round-mode=places, round-precision=0]{0.0} & \num{6.554e-45} & \num{0.178} & \num{1.563e-18} & \ce{N,C,O} & This work\\
            & & & & \num{2.979e-46} & & & Others & This work\\
            7f & D & \ce{O2 + M -> O + O + M} & \num{8.249e-19} & \num{1.661e-8} & \num[round-mode=places, round-precision=1]{-1.5} & \num{8.249e-19} & \ce{N,C,O} & \citep{park1993review}\\
            & & & & \num{3.321e-9} & & & Others & \citep{park1993review}\\
            7b & R & \ce{O + O + M -> O2 + M} & \num[round-mode=places, round-precision=0]{0.0} & \num{9.655e-42} & \num{-0.906} & \num{8.249e-19} & \ce{N,C,O} & This work\\
            & & & & \num{1.931e-42} & & & Others & This work\\
		\end{tabular}}
    \label{TCErates_dissrecomb}
\end{table*}

\begin{table*}
	\centering
	\caption{Chemical kinetics rates for exchange reactions used in the TCE model. $E_{a}$ represents the Arrhenius activation energy, $A$ the Arrhenius pre-exponential factor, $n$ the modified Arrhenius temperature exponent and $\Delta H$ the reaction enthalpy. The $type$ represents Exchange (E) reactions (see \citealp{ibragimova1991recommended,park1994review,ref:johnston2014,gokcen2007n2,fujita2006impact,park1993review}).}
		\resizebox{\textwidth}{!}{\begin{tabular}{lllrrrrll}
			\hline
            \hline
			i & Type & Reaction & E$_a$ (J) & A (\unit{m^{3}.molecule^{-1}.s^{-1}}) & n & $\Delta$H (J) & Ref. \\
            \hline
            8f & E & \ce{CO2 + O -> O2 + CO} & \num{4.667e-19} & \num{4.500e-16} & \num[round-mode=places, round-precision=0]{0.0} & \num{-5.614e-20} & \citep{ibragimova1991recommended}\\
            8b & E & \ce{O2 + CO -> CO2 + O} & \num{4.102e-19} & \num{1.616e-17} & \num{0.194} & \num{5.614e-20} & This work\\
            9f & E & \ce{CO + C -> C2 + O} & \num{8.008e-19} & \num{3.985e-13} & \num[round-mode=places, round-precision=1]{-1.0} & \num{-8.008e-19} & \citep{park1994review}\\
            9b & E & \ce{C2 + O -> CO + C} & \num{1.142e-20} & \num{1.949e-12} & \num{-1.405} & \num{8.008e-19} & This work\\
            10f & E & \ce{CO + N -> CN + O} & \num{5.329e-19} & \num{1.827e-15} & \num[round-mode=places, round-precision=0]{0.0} & \num{-5.329e-19} & \citep{ref:johnston2014}\\
            10b & E & \ce{CN + O -> CO + N} & \num[round-mode=places, round-precision=0]{0.0} & \num{5.203e-15} & \num{-0.320} & \num{5.329e-19} & This work\\
            11f & E & \ce{CO + NO -> CO2 + N} & \num{1.836e-19} & \num{3.576e-21} & \num[round-mode=places, round-precision=1]{0.2} & \num{-1.836e-19} & This work\\
            11b & E & \ce{CO2 + N -> CO + NO} & \num{5.944e-20} & \num{4.539e-16} & \num{-0.787} & \num{1.836e-19} & This work\\
            12f & E & \ce{CO + O -> O2 + C} & \num{9.554e-19} & \num{6.476e-17} & \num[round-mode=places, round-precision=2]{-0.18} & \num{-9.554e-19} & \citep{park1994review}\\
            12b & E & \ce{O2 + C -> CO + O} & \num[round-mode=places, round-precision=0]{0.0} & \num{3.537e-16} & \num{-0.472} & \num{9.554e-19} & This work\\
            13f & E & \ce{C2 + N2 -> CN + CN} & \num{2.899e-19} & \num{2.491e-17} & \num[round-mode=places, round-precision=0]{0.0} & \num{-2.899e-19} & \citep{gokcen2007n2}\\
            13b & E & \ce{CN + CN -> C2 + N2} & \num{1.390e-19} & \num{2.648e-17} & \num{0.036} & \num{2.899e-19} & This work\\
            14f & E & \ce{CN + C -> C2 + N} & \num{2.499e-19} & \num{4.982e-16} & \num[round-mode=places, round-precision=0]{0.0} & \num{-2.499e-19} & \citep{fujita2006impact}\\
            14b & E & \ce{C2 + N -> CN + C} & \num{1.694e-21} & \num{1.805e-15} & \num{-0.155} & \num{2.499e-19} & This work\\
            15f & E & \ce{CN + O -> NO + C} & \num{2.016e-19} & \num{2.657e-18} & \num[round-mode=places, round-precision=1]{0.1} & \num{-2.016e-19} & \citep{ref:johnston2014}\\
            15b & E & \ce{NO + C -> CN + O} & \num[round-mode=places, round-precision=0]{0.0} & \num{8.677e-19} & \num[round-mode=places, round-precision=1]{0.2} & \num{2.016e-19} & This work\\
            16f & E & \ce{N + CO -> NO + C} & \num{7.386e-19} & \num{1.823e-16} & \num[round-mode=places, round-precision=2]{0.07} & \num{-7.386e-19} & \citep{fujita2006impact}\\
            16b & E & \ce{NO + C -> N + CO} & \num[round-mode=places, round-precision=0]{0.0} & \num{9.856e-18} & \num[round-mode=places, round-precision=2]{0.15} & \num{7.386e-19} & This work\\
            17f & E & \ce{N2 + C -> CN + N} & \num{3.203e-19} & \num{1.823e-16} & \num[round-mode=places, round-precision=2]{-0.11} & \num{-3.203e-19} & \citep{park1994review}\\
            17b & E & \ce{CN + N -> N2 + C} & \num{2.556e-20} & \num{2.118e-13} & \num[round-mode=places, round-precision=2]{-1.01} & \num{3.203e-19} & This work\\
            18f & E & \ce{N2 + CO -> CN + CO} & \num{1.063e-18} & \num{1.933e-14} & \num[round-mode=places, round-precision=2]{-1.23} & \num{-1.063e-18} & \citep{fujita2006impact}\\
            18b & E & \ce{CN + CO -> N2 + CO} & \num{1.875e-20} & \num{5.792e-13} & \num{-1.983} & \num{1.063e-18} & This work\\
            19f & E & \ce{N2 + O -> NO + N} & \num{5.246e-19} & \num{9.963e-17} & \num[round-mode=places, round-precision=1]{0.1} & \num{-5.246e-19} & \citep{fujita2006impact}\\
            19b & E & \ce{NO + N -> N2 + O} & \num{1.524e-20} & \num{7.190e-16} & \num{-0.288} & \num{5.246e-19} & This work\\
            20f & E & \ce{NO + O -> O2 + N} & \num{2.684e-19} & \num{1.389e-17} & \num[round-mode=places, round-precision=0]{0.0} & \num{-2.684e-19} & \citep{park1993review}\\
            20b & E & \ce{O2 + N -> NO + O} & \num[round-mode=places, round-precision=0]{0.0} & \num{4.601e-15} & \num{-0.546} & \num{2.684e-19} & \citep{park1993review}\\
			\hline
		\end{tabular}}
    \label{TCErates_exchange}
\end{table*}  
\section{Maximum Grid Refinement Level} 
\label{sec:AMR}

Simulations were also performed to study the effect of the maximum adaptive mesh refinement (AMR) level. For this particular study, we focused on three noble gas isotopes, $^{4}$He, $^{40}$Ar and $^{132}$Xe, which represent a light, medium mass, and heavy isotope, respectively. For each noble gas, two separate simulations were conducted, in which the maximum AMR level was capped to 8 and 9, respectively. Otherwise, the baseline conditions were used for these simulations. Table~\ref{tab:maxAMR} shows the results for the reference simulations (where the AMR level is capped at 10), as well the simulations with that value capped at 8 and 9. It can be seen that while the absolute relative error between maximum AMR 10 and 9 is relatively low (under 2\,\%), that absolute error then increases between when comparing AMR 10 and 8 (up to 7\,\%). Therefore, while using a maximum AMR of 10 might not be entirely satisfying with respect to resolving a fraction of the mean free path in the regions of the highest density, the results show that the error in the normalized mass fraction decreases as the maximum AMR approaches 10.

\begin{table}
    \centering
    \caption{Normalized mass fraction for three noble gas species, at baseline freestream conditions, as a function of the maximum adaptive mesh refinement (AMR) level.}
    \resizebox{0.4\columnwidth}{!}{\begin{tabular}{cccc}
            \hline\hline
            Noble gas  & $^{4}$He & $^{40}$Ar & $^{132}$Xe \\
            \hline
            AMR 10 (ref.)   & \num{0.25766}  & \num{1.41831}   & \num{2.11281}    \\
            \hline
            AMR 9                & \num{0.26149}  & \num{1.41350}   & \num{2.07067}    \\
            Relative difference  & \num[round-mode=places, round-precision=2]{1.488}\,\%  & \num[round-mode=places, round-precision=2]{-0.339}\,\%  & \num[round-mode=places, round-precision=2]{-1.994}\,\%   \\
            \hline
            AMR 8                & \num{0.23985}  & \num{1.46113}   & \num{2.25405}    \\
            Relative difference  & \num[round-mode=places, round-precision=2]{-6.912}\,\% & \num[round-mode=places, round-precision=2]{3.019}\,\%   & \num[round-mode=places, round-precision=2]{6.685}\,\%    \\
        \end{tabular}}
    \label{tab:maxAMR}
\end{table}
\section{Fractionation of Noble Gases}
\label{sec:fractionation}

\subsection{Theory}
\label{sec:theory}
In his original DSMC manuscript~\citep{bird1994molecular} discusses the fractionation of noble gases in the context of several thermodynamic phenomena. First, in Section 3.5, he discusses the thermal and pressure diffusion of gases: \say{thermal diffusion causes the larger and generally heavier molecules to migrate towards the cooler regions of a flow}, and \say{pressure diffusion causes the heavy and light gases to move towards regions of high and low pressure, respectively}. Additionally, \say{pressure diffusion and thermal diffusion act in the same direction in the stagnation region of the hypersonic flow over a cold blunt body and the separation can be significant in the flow of a partly dissociated gas which necessarily involves large mass ratios.}

In section 12.7, Bird validates thermal diffusion with the DSMC methodology. Subsequently, in Section 12.11, he investigates normal shock waves, specifically for a helium--xenon mixture, with xenon as the trace species (3\,\% mole fraction). Confirming previously observed experimental results~\citep{gmurczyk1979shock}, Bird found that \say{normal shocks exhibited the leading of the heavy gas profile by the light gas profile,} referring to density profiles. Additionally, Bird's DSMC simulations show that the \say{xenon concentration at the center of the shock drops to less than half the upstream and downstream values.} From the mole fraction plot, it is also noticeable that the upstream value of the mole fraction is recovered downstream, i.e. the mole fraction of Xe is the same upstream and downstream of the shock. The diffusion velocity plots also show that the heavy species, \ce{Xe}, diffuses faster than the light species, \ce{He}.

In Section 14.7, Bird then explores diffusion in blunt-body flows by looking at a helium--xenon mixture (71.78\,\% \ce{He} by mole) impinging on a vertical flat plate. He finds that \ce{Xe} is concentrated in the stagnation region, and that in contrast, the wake of the flow is composed almost entirely of \ce{He}. Bird also uses these various sections to expand on the fact that the Navier-Stokes equation inadequately represent diffusion by neglecting the thermal and pressure diffusion terms, though it has recently been demonstrated that it is possible to include these terms in continuum CFD solvers~\citep{kroells2025investigation}.

\begin{figure*}
	\begin{subfigure}{.32\textwidth} 
        \centering
        \includegraphics[width=\linewidth]{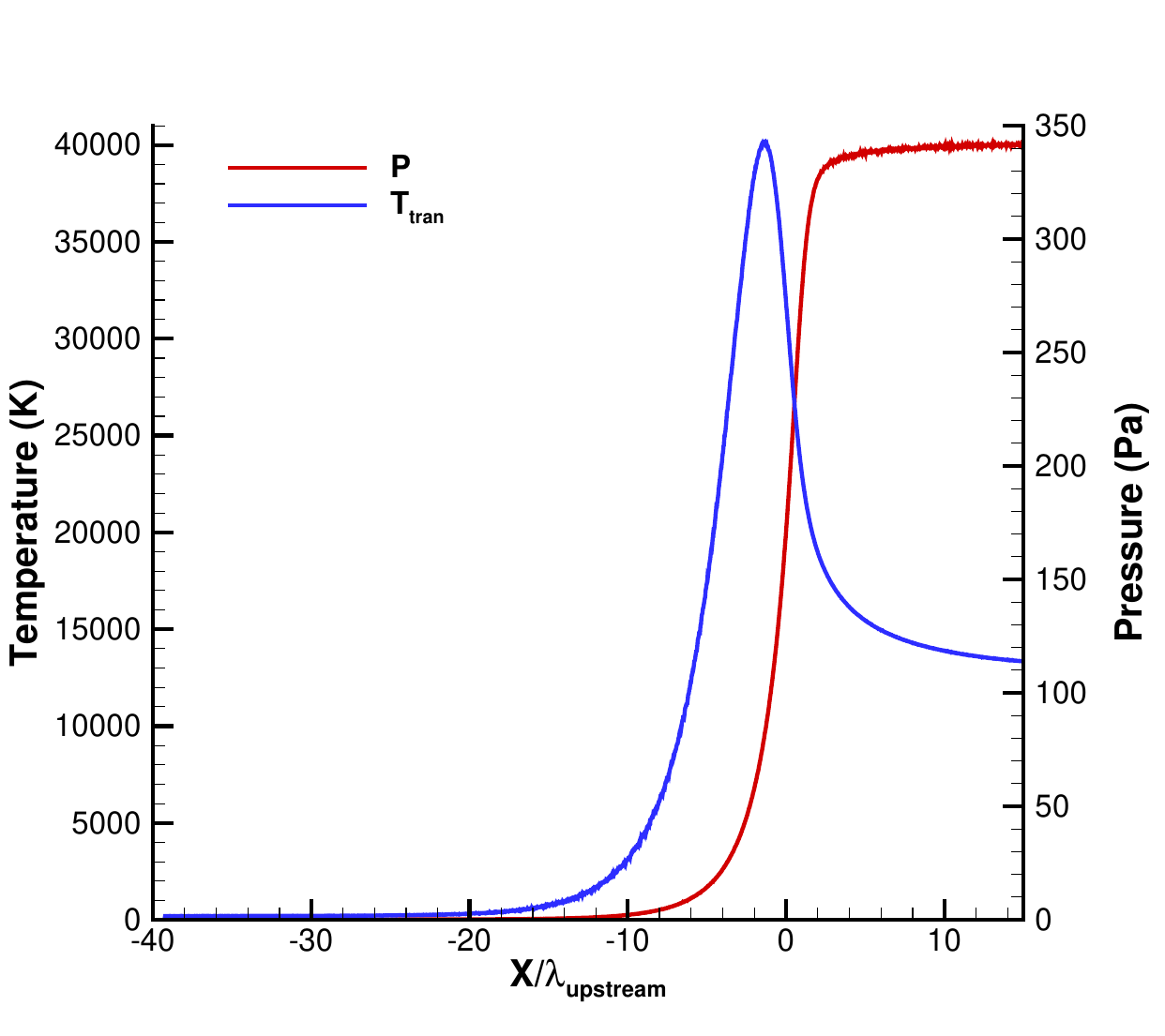}
        \caption{Temperature and pressure.} \label{fig:T-and-P-He-shock}
	\end{subfigure}
    \hfill
	\begin{subfigure}{.32\textwidth}
		\centering
        \includegraphics[width=\linewidth]{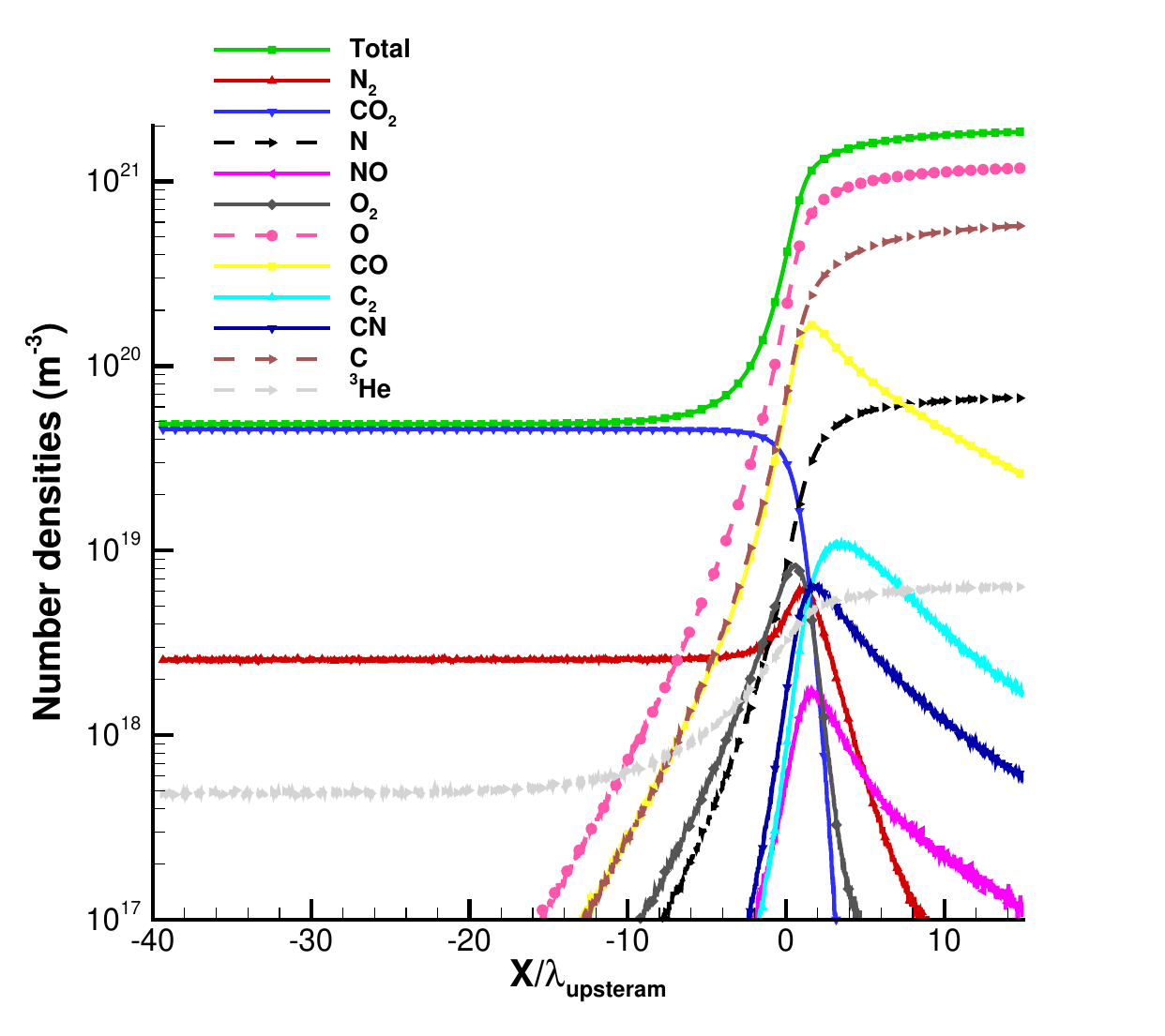}
		\caption{Number densities.}  \label{fig:Numdens-He-shock}
	\end{subfigure}
    \hfill
	\begin{subfigure}{.32\textwidth}
		\centering
        \includegraphics[width=\linewidth]{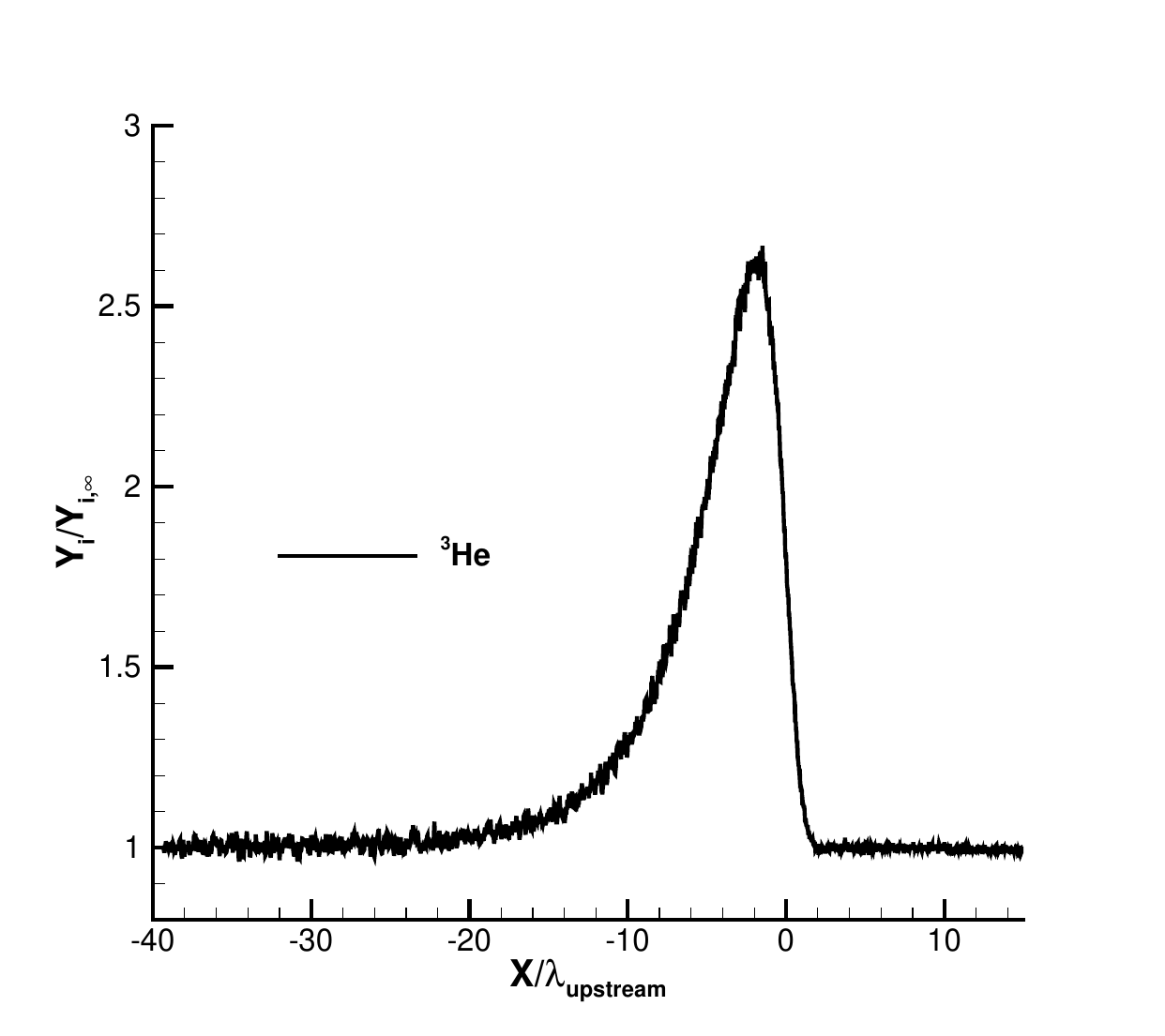}
		\caption{Normalized $^{3}$\ce{He} mass fraction.}  \label{fig:Y3He-He-shock}
	\end{subfigure}
 		\caption{Distribution of various quantities of interest in a Mach 49.5 shock with composition equivalent to that of Venus at 110~km altitude. The only noble gas modeled is $^{3}$\ce{He}.}  \label{fig:1D-shock-plots}
 \end{figure*}

\subsection{Verification}
\label{sec:verification}
We replicate some of Bird's normal shock simulations using the freestream conditions of the Venusian atmosphere relevant to this mission concept. We perform 1-dimensional normal shock simulations for the conditions listed in Table~\ref{tab:sim-matrix}. The simulation setup is similar to that depicted in~\citep[Ch. 2.2.2]{higdon2018monte} and~\citep{zhu2014modeling}, for unsteady hypersonic shocks, given that the post-shock conditions are unknown. We model unsteady shocks in SPARTA by initializing a finite 1-dimensional domain (4000 cells in the streamwise direction, 1 cell in each of the two other directions, with periodic boundary conditions applied in those directions). One end of the domain is open to an inlet where particles sampled from a Maxwellian velocity distribution are added at every time step. The other end of the domain is closed with a specular wall. A shock reflects from the specular end wall and propagates upstream. Because the shock speed is not known \textit{a priori}, an iterative method is used, where the reflected shock velocity is computed as the shock propagates through the domain, and the input velocity is adjusted to converge towards the desired velocity of 10.5~km/s. To reduce statistical noise, we perform sampling over a spatial region downstream of the reflected shock and also use a moving window technique (\emph{i.e.}, the frame of reference is redefined with the center of shock as its origin). Finally, macro parameters such as temperatures, pressure, and number densities are sampled and averaged.
We also use the Mutation$^{++}$~\citep{Scoggins2020} software to compute post-normal shock equilibrium conditions for our conditions of interest. Given freestream conditions (pressure, temperature and velocity), Mutation$^{++}$ computes the Rankine-Hugoniot relations and the relaxed, thermodynamic and chemical equilibrium state, and velocity.

We perform independent 1-dimensional shock verification simulations for three noble gases of interest to this study: the lightest noble gas, $^{3}$\ce{He}, the heaviest one, $^{136}$\ce{Xe}, and one of mass similar to that of the mixture, $^{40}$\ce{Ar}. For brevity, the results shown thereafter are those for one noble gas, $^{3}$\ce{He}. Fig.~\ref{fig:T-and-P-He-shock} shows the translational temperature and pressure across the normal shock as a function of the distance normalized by the mean free path upstream of the shock. First, one can see that at those conditions, the shock is quite diffuse and spans over 10 mean free paths. While the pressure upstream of the shock is 0.13~Pa, it rises to 341~Pa downstream, indicating a pressure ratio of \num{2623}. Mutation$^{++}$ reports a post-shock, equilibrium pressure of 355.9~Pa, which is in excellent agreement with our DSMC calculations using finite-rate chemistry.
The translational temperature peaks at \num{40000}~K, and reaches a plateau of $\approx$\num{12800}~K in the downstream region away from the wall. The number densities of the different gases across the shock can be seen in Fig.~\ref{fig:Numdens-He-shock}. As expected at Mach $\approx50$ conditions, the \ce{CO2} in the freestream gets entirely dissociated into \ce{CO}, \ce{O} and \ce{C}, with the atomic species predominant downstream. The \ce{N2} also gets entirely dissociated into \ce{N} atoms. Finally, in Fig.~\ref{fig:Y3He-He-shock}, the normalized $^{3}$\ce{He} mass fraction is plotted with respect to the normalized shock length. 
The $^{3}$\ce{He} in the center of the shock increases to over 2.5 times its upstream and downstream value, but recovers to the same value downstream as upstream, as discussed in the previous section. While not shown in the plots, $^{136}$\ce{Xe} shows the opposite trend, and the value in the center of the shock drops to 20\,\% of the downstream value, and $^{40}$\ce{Ar} drops to 60\,\% of its downstream value.

\printcredits

\bibliographystyle{cas-model2-names}

\bibliography{references}



\end{document}